\documentclass[twoside,11pt,theapa]{article}
\usepackage{jair, theapa, rawfonts}

\usepackage[utf8]{inputenc}

\usepackage{amsthm}
\usepackage{graphicx}
\usepackage{natbib}
\usepackage{color}
\usepackage{xcolor}
\usepackage{url}
\usepackage{dblfloatfix}
\usepackage{hhline}
\usepackage{colortbl}
\usepackage{diagbox}
\usepackage{multirow}
\usepackage{multicol,lipsum}
\usepackage{amsmath}
\usepackage{amssymb}
\usepackage[mathscr]{euscript}
\usepackage{tikz}
\usepackage{soul}
\usepackage{pgfplots}
\usepackage{gensymb}
\usepackage{verbatim}
\usepackage{algorithm}
\usepackage{algpseudocode}
\usepackage{adjustbox}
\usepackage{subcaption}
\usepackage{adjustbox}
\usepackage[toc,page]{appendix}
\usepackage{longtable}
\usepackage{mwe}
\usepackage[algo2e,ruled,linesnumbered,lined]{algorithm2e}
\usepackage{changes}
\usepackage{stmaryrd}
\usepackage{hyperref} 

\usepackage[nounderscore]{syntax} 

\def\CHANGES

\AtBeginDocument{%
  \providecommand\BibTeX{{%
    \normalfont B\kern-0.5em{\scshape i\kern-0.25em b}\kern-0.8em\TeX}}}
    
\definecolor{Light_Gray}{HTML}{EFEFEF}
\definecolor{Light_Orange}{HTML}{F9CB9C}
\definecolor{Light_Blue}{HTML}{9FC5E8}
\definecolor{Light_Green}{HTML}{D9EAD3}
\definecolor{Light_Yellow}{HTML}{FCE5CD}
\definecolor{Gray}{gray}{0.9}

\newtheorem{definition}{Definition}

\newtheorem{example}{Example}

\newtheorem{proposition}{Proposition}

\newcommand\pfun{\mathrel{\ooalign{\hfil$\mapstochar\mkern5mu$\hfil\cr$\to$\cr}}}

\graphicspath{{./Figures/}}
\graphicspath{{./Figures/example}}
\graphicspath{{./Figures/experiment}}

\newcommand{\nocontentsline}[3]{}
\newcommand{\tocless}[2]{\bgroup\let\addcontentsline=\nocontentsline#1{#2}\egroup}

\let\oldnl\nl
\newcommand{\nonl}{\renewcommand{\nl}{\let\nl\oldnl}}

\pgfplotsset{compat=1.18}
\begin{document}

\title{Causality for Cyber-Physical Systems}






\author{\name Hugo Araujo \email hugo.araujo@kcl.ac.uk \\
       \name Hana Chockler \email hana.chockler@kcl.ac.uk \\
       \name Mohammad Reza Mousavi \email mohammad.mousavi@kcl.ac.uk \\
       \addr King's College London
       \AND
       \name Gustavo Carvalho \email ghpc@cin.ufpe.br \\
       \name Augusto Sampaio \email acas@cin.ufpe.br \\
       \addr Universidade Federal de Pernambuco}

\maketitle

\begin{abstract}
  We present a formal theory for analysing causality in cyber-physical systems. To this end,  we extend the theory of actual causality by Halpern and Pearl to cope with the continuous nature of cyber-physical systems. Based on our theory, we develop an analysis technique that is used to uncover the causes for examples of failures resulting from verification, which are represented as continuous trajectories. We develop a search-based technique to efficiently produce such causes and provide an implementation for such a technique. Moreover, we apply our solution to case studies (a suspension system and a connected platoon) and benchmark systems to evaluate its effectiveness; in the experiment, we show that we were able to detect causes for inserted faults. 
\end{abstract}


\section{Introduction} 
\label{sec:introduction}


Cyber-physical systems (CPSs) are systems that integrate computation with physical processes, in contexts where communication networks and human interaction may be present. Embedded computers and networked embedded components in CPSs monitor and control the physical processes, usually with feedback loops where physical processes affect computations and vice versa \cite{lee2016introduction}. In CPSs, components often operate in both spatial and temporal dimensions, where there is an intense link between physical and computational elements.


The importance of reliability in safety-critical CPSs warrants further research into their verification. Particularly, there is a need to devise verification techniques based on mathematical relations, e.g., notions of conformance~\cite{roehm2019model}, that decide whether the system behaves as expected. Conformance notions are employed by conformance testing processes to evaluate whether a system behaves as required by its specification. Conformance testing and falsification approaches \cite{tretmans1996conformance, abbas2014conformance, van2006hybrid, weiglhofer2009fault, roehm2019model, annpureddy2011s, donze2010breach, zhang2018two, mitsch2016modelplex} have the benefit of offering mathematical assurances about the correctness of the system.

Although they are undoubtedly useful, verification techniques can be of limited use, particularly in the context of complex systems, without a systematic way to trace the failures back to their original causes. Testing a system without a method for locating the source of the discovered faults can lead to an incredible amount of resources spent on manual system inspections.

In this context, causal analysis is an essential ingredient of such rigorous verification techniques, providing effective means to isolate and eventually remove the faults causing hazards and failures observed in the verification process. There have been many attempts to define and analyse causality. In a philosophical context, causality is defined the relationship between events where the the cause contributes to the realisation of a different event (the effect), where the cause is (partially) responsible for the effect, which is (partially) dependent on its cause \cite{zalta1995stanford}. In a more practical context, Halpern and Pearl have made use of structural equation models to provide a mathematical definition of actual causality \cite{halpern2005causes, halpern2005causes2}.  Chockler builds upon this work by integrating the notion of actual cause into a rigorous verification process \cite{chockler2004responsibility}. 

However, most definitions of actual causality only cater for discrete systems and do not cover the complexity of dynamic physical phenomena. Due to the continuous and quantitative nature of the physical parts of CPSs, the notion of causality is bound to have a quantitative nature and accommodate physical dynamics.

%

\subsection{Problem definition and contributions} 

Our research goal is to propose a formal theory and a practical analysis technique for checking causality that can cope with the complexity of CPSs. The main contributions of this work are the following:

\begin{itemize}
    \item We extend the theory of actual causality \cite{halpern2005causes} to cope with CPSs. The new theory takes time and continuous dynamics into consideration. 
    
    \item We develop a process for checking causality and mechanise the associated algorithms in MATLAB and integrate them with our tool, HyConf \cite{araujo2019multiobjective}, for conformance testing of CPSs.

    \item We propose and discuss practical applications of causality in the verification of CPSs.
    
    \item We apply the developed algorithms and tools to a range of case studies and evaluate the effectiveness of our strategy.
\end{itemize}

Applying a systematic and formal methodology to determine the cause of a failure in a CPS requires structured notions of faults and causes using mathematical notations. We work in a mathematical framework where faults are represented using first-order formulae, in which atomic predicates are expressed over signal values (akin to safety subset of Signal Temporal Logic~\cite{maler2004monitoring}). Causes are expressed using intervals of trajectories, which we call \emph{trajectory slices}, of the system variables leading to the falsification of such properties.


Causes are determined using a causal model, which is a representation of the system dynamics using structural equations. In this model, we separate variables into endogenous and exogenous sets; the former represents the set of variables that are chosen to undergo the causal analysis and can be seen as potential causes. The latter are variables that affect the system but are not seen as potential causes. In a practical setting, having too many endogenous variables greatly increases the costs of the causal analysis in CPSs. Hence, the choice of endogenous and exogenous variables is left to the user and has an impact on the variables that can appear in the cause and the efficiency of the causal analysis.

Furthermore, for a notion of causality for CPSs, one needs to take time into consideration. For instance, faults may have been caused by trajectory slices occurring at specific time intervals. That is, if these trajectory slices had occurred in different intervals, the fault may not have happened. Thus, one cannot ignore the history of the system execution. This way, we aim to identify not only to which components the fault should be attributed, but also provide the intervals of time when the causal behaviour has occurred in the respective components.

Finally, one must consider the continuous and infinite nature of physical phenomena. The way system variables affect each other can determine whether a cause comprises one or multiple interacting variables. It is possible that multiple alternative causes can be found for a single failure. 

We mechanised our causal analysis using the Matlab/Simulink framework, which is a commonly used environment for modelling and analysis of control systems, thus increasing the accessibility of our strategy to the CPSs community. We developed a search-based algorithm to find the causes and integrated the mechanisation of our theory of actual causality into our pre-existing tool for testing cyber-physical systems \cite{araujo2019multiobjective}. We applied the developed algorithms and tools to a range of case studies to demonstrate the effectiveness of our strategy, where we pinpoint causes and present explanations for failures that have been detected in the verification process.



\subsection{Structure of the paper}

This paper is organised as follows. In Section \ref{sec:relatedWork}, we present the related work. In Section \ref{sec:preliminaries}, we discuss the preliminaries used to develop our framework. In Section \ref{sec:theoryContinuous}, we present our extension to the theory of  causality that considers time and continuous dynamics and, in Section \ref{sec:applications}, we develop the mechanisation of this extended theory. In Section \ref{sec:caseStudies}, we design and conduct experiments to show how our strategy can be applied to complex systems. Finally, in Section \ref{sec:conclusion}, we draw some conclusions and point out the directions of our future research.

\section{Related work}
\label{sec:relatedWork}

Our context is the theory of actual causality \cite{halpern2001causes} where, in a given scenario leading to an outcome, the events are analysed in order to find causes. This is also called token-level causality, which concerns causal relations regarding particular events and settings. It is in contrast with type-level causality \cite{hausman2005causal} where general causal rules governing a system are sought. 

There are a few variations of the definition of actual causality. The original one is by Halpern and Pearl \cite{halpern2001causes}, which is followed by two variants (updated and modified) by Halpern and Pearl, and Halpern alone, respectively \cite{ halpern2005causes, halpern2015modification}; our study is based on the definition by Halpern and Pearl in 2015 \cite{halpern2015modification}, which is called the \emph{modified} definition. In the most recent version of the theory \cite{halpern2015modification}, Halpern simplifies the impact of contingencies, which can be seen as pre-conditions for a cause but not part of the cause itself. More specifically, the key difference between the \emph{modified} definition and the other definitions is the requirement that the contingencies in the modified version should be set to their initial values, whereas in the updated definition, contingencies can be set to different values. 

Halpern and Pearl also introduced the notion of time-indexed endogenous variables~\cite{halpern2005causes}; this treatment of time in the original theory of Halpern is discrete in nature and does not cater for continuous time and dynamics that are necessary to model cyber-physical systems. In order to use this definition of Halpern and Pearl, one needs to define a fixed discretisation of continuous variables and come up with a causal model explaining their relationships. Our theory, however,  provides an abstraction layer that works directly with the continuous specification of the dynamics and builds the necessary sampling as a part of the analysis.  In a recent extension of their work, Peters and Halpern propose the generalised structural equations models (GSEM) \cite{peters2021causal} where a signature comprises a set of interventions and the equations map an intervention to  a set of outcomes (e.g., when variable X is set to x, then variable Y is equal to y and variable Z is equal to z). It brings about the possibility of having an infinite set of valuations for variables. 
Furthermore, Halpern shows that this can be adapted to ordinary differential equations; in his treatment, instead of having a set of endogenous variables,  the signature comprises the value of a variable at any moment in time, as defined by differential equations. 
This recent extension may be a theoretical alternative to our proposed framework; however, to apply any practical causal analysis on this infinite set of variables, further abstractions and algorithmic procedures need to be developed. In principle, our work can be adapted to deal with the many variants of Halpern and Pearl's theory of actual causality. 

Apart from the variants developed by Halpern and Pearl, there are many other definitions of actual causality by others. In the remainder of this section, we review some of the most relevant variants or applications of actual causality  developed by other researchers. Subsequently, we also mention some alternative theories, to actual causality, that can be used and extended for finding root causes in CPSs.

Several works employ the definitions of causality by Halpern and Pearl to formal verification; often as a reasoning tool for explaining counterexamples in the discrete domain. The most relevant to our work are discussed below. 

Baier et al.\ \cite{Baier2021From} have conducted a survey on published approaches that utilise Halpern-Pearls's notion of causality. More precisely, they look into formal approaches to probabilistic causation that can explain observable behaviour in reactive systems.

Within the context of cyber-physical systems, Deng et al.\ ~\cite{deng2023causal} have proposed a temporal logic for analysing causality called causal temporal logic. Once the prospective causes and effects are expressed in this temporal logic, they assess causality by generating traces that can satisfy/violate the causal formula and, thus, calculating the degree of sufficiency and necessity. There are key differences between our work and theirs. Firstly, we propose a conservative extension of Halpern-Pearls theory, in which we aim to keep the nuances of their theory (such as the notions of contingency and causal path, which are key factors to consider when determining actual causality) and not just the notions of sufficiency and necessity. Furthermore, their work assumes access to the correct behaviour and prospective causes. In our work, we do not need access to the correct behaviour; we search for causes by making use of meta-heuristics and looking for behaviours that can violate/satisfy the effect (formula).

Leitner-Fischer and Leue \cite{leitner2013causality} define a theory of causality that considers the temporal order as well as the non-occurrence of events. They also provide a search-based on-the-fly causality assessment that does not require the counterexamples to be generated in advance. There, even though the order of events is important, no concrete notion of time is introduced. In our work, however, real time plays an important role and trajectories are represented in the continuous time domain. 

Caltais, Mousavi, and Singh \cite{caltais2020causal} define a theory of actual causality for labelled transition systems. Their formalisation is inspired by the definition of Halpern and Pearl. Their main result is a theory to explain counterexamples in model checking with respect to properties in Hennesy-Milner Logic \cite{hennessy1980observing}. They mechanise their theory in a prototype tool, which  interacts with the mCRL2 model-checker \cite{groote2014modeling} in order to check the various conditions in their definition of actual causality. Our work shares a similar nature of employing causality to interpret traces leading to failures. However, we employ models and logic that consider quantitative aspects, such as trajectories that are solutions to systems of differential equations.

Ibrahim et al.\ \cite{ibrahim2019practical} provide a process to convert attack trees, fault trees, and timed failure propagation graphs for CPSs into Halpern-Pearl causal models. They illustrate their approach using an Unmanned Aerial Vehicle case study. Even though their work, like ours, focuses on CPS-related aspects, it does not handle continuous aspects of such systems.

In the context of hardware verification, Chockler, Grumberg, and Yadgar \cite{chockler2008efficient} employ a notion of responsibility (degree of causality) \cite{chockler2004responsibility} to improve the quality of abstraction refinement by producing mode efficient counterexamples. Besides the continuous aspects, our approach incorporates the modelling of platform (hardware), controllers (software) and environment into a single model that considers a high-level abstraction of the system. We do not consider such a notion of responsibility, however; this is one of the directions for our future work.

There are quantitative extensions of the theory of actual causality. Pearl studies the effect of causality in probabilistic systems \cite{pearl1995causal, pearl2014probabilistic}, providing the underlying theory for causal inference \cite{spirtes2010introduction, pearl2009causal}, which provides the mathematical tools and the language for articulating probabilities of causation. His work employs structural equations to cope with counterfactuals and randomisation and has applications in AI \cite{pearl2019seven}. Baier et al.\ \cite{baier2022probability, Baier2021From} introduce and formalise cause-effect relation in Markov decision processes using the probability-raising principle. They provide algorithms for checking cause-effect relationships and the existence of probability-raising causes for given effect scenarios. To our knowledge their work does not use counter-factual reasoning and is not formally related to our formal theory of actual causality. 
With respect to the use of causal analysis in the formal verification process itself (and not just as a way to explain the results), Baier et al. \cite{baier2022operational} have presented a temporal logic characterisation for the notions of sufficiency and necessity, which are based on the concepts found in causal reasoning. They propose an optimisation algorithm for the computation of causes based on these degrees of necessity and sufficiency. Unlike our work, however, they do not cater for the continuous properties of physical systems but they do however consider stochastic aspects.
 We consider extending our results to quantitative and probabilistic notions as a worthwhile future direction. 

Zhang et al. \cite{zhang2023online} propose a method for the online monitoring of Signal Temporal Logic (STL). Instead of computing the distance of the system's output against the specification (called  robustness value~\cite{fainekos2009robustness}, also in our approach), they compute whether an "instant" is relevant to a violation and how far the instant is from a violation. This bears some resemblence to our notion of actual cause. However, their work, unlike ours, does not formally relate to a theory of actual causality and does not pinpoint a cause by tying specific variables to specific time intervals.

Beer et al.\ \cite{beer2009explaining} use actual causality to explain counterexamples in hardware verification for Linear Temporal Logic properties. The proposed algorithm is implemented in the IBM RyleBase PE tool, where causality is applied to traces but ignores the system model from which the traces originated. Our assessment of causality considers both the setting and the model, which in our case also encompasses continuous and discrete aspects. This richer setting leads to a more computationally-intensive analysis but allows us to find causes in more complex systems and with more precision (steered by the model).

Dubslaff et al.\ \cite{dubslaff2022causality} use counterfactual reasoning to identify causes in configurable systems. This is done by identifying the features and interactions that are the reason for emerging functional and non-functional properties. They call this concept feature causality. These notions are in clear contrast with the notion employed in our work where we start with a concrete scenario leading to the effect. Our choice is justified by our context, where we employ causal analysis as a step in the testing and verification process, where finding an error-trace or counterexample initiates the causal analysis process. 

There are other theories of causality that are not concerned with a particular scenario. For example, Granger's causality \cite{granger1969investigating} is a statistical concept that checks the possibility of a time series predicting another. It employs notions of trends, seasonal patterns and noise in time series forecasting and can be applied to machine learning, finance and weather forecast \cite{shojaie2022granger}.

Table \ref{tab:relatedwork} summarises the key difference between the above-surveyed techniques and our proposed framework by showing the supported features for each technique.

\begin{table}[!h]
\centering
\caption{A comparison between different methods for causal analysis.}
\label{tab:relatedwork}
\resizebox{1\textwidth}{!}{%
\begin{tabular}{l|c|c|c|c|c|c|}
\cline{2-7}
 & \multicolumn{1}{c|}{\textbf{Formal Models}} & \textbf{Counterexamples} & \textbf{Hybrid Systems} &  \textbf{Mechanisation} & \textbf{Search-based} & \textbf{Stochastic} \\ \hline

\multicolumn{1}{|l|}{Ours} & \checkmark & \checkmark & \checkmark & \checkmark & \checkmark & \\ \hline
\multicolumn{1}{|l|}{Leitner-Fischer \cite{leitner2013causality}} & & & & \checkmark & \checkmark & \\ \hline
\multicolumn{1}{|l|}{Caltais \cite{caltais2020causal}} & \checkmark & \checkmark & & \checkmark & & \\ \hline
\multicolumn{1}{|l|}{Chockler \cite{chockler2008efficient}} & & \checkmark &  & \checkmark &  & \\ \hline
\multicolumn{1}{|l|}{Beer \cite{beer2009explaining}} & \checkmark & \checkmark & & \checkmark & &  \\ \hline
\multicolumn{1}{|l|}{Pearl \cite{pearl2019seven}} & \checkmark & \checkmark &  &  &  & \checkmark \\ \hline
\multicolumn{1}{|l|}{Granger \cite{granger1969investigating}} & \checkmark &  &  & \checkmark &  & \\ \hline
\multicolumn{1}{|l|}{Deng ~\cite{deng2023causal}} & \checkmark & & \checkmark & \checkmark & \checkmark & \\ \hline
\multicolumn{1}{|l|}{Zhang\cite{zhang2023online}} & \checkmark & & \checkmark & \checkmark & & \\ \hline
\multicolumn{1}{|l|}{Peters \cite{peters2021causal}} & \checkmark & \checkmark & \checkmark &  &  & \\ \hline
\multicolumn{1}{|l|}{Baier \cite{baier2022probability, Baier2021From}} & \checkmark & \checkmark &  & \checkmark &  & \checkmark \\ 
\hline

\end{tabular}%
}
\end{table}


There are fault localisation techniques  that are not causality-based. For instance, fault-tree analysis \cite{ericson1999fault} is an established method to investigate faults in safety critical systems. In fault trees, a graphical representation captures the logical connections between faults and their origins. Similar to our causal analysis, fault tree analysis starts from a failure event, which is represented at the top of the tree and then it is worked backwards to determine the root causes. In order to cope with dynamic systems, fault trees have been extended to include quantitative dependability analysis. Several approaches have been proposed and used such as dynamic fault trees \cite{dugan1992dynamic}, state-event fault trees \cite{kaiser2007state}, spectrum-based fault localisation \cite{abreu2009practical}, and Stochastic Hybrid Fault Tree Automaton \cite{chiacchio2016shyfta}. Temporal Fault Trees \cite{PALSHIKAR2002137} are one such extension, which, to our knowledge, can only handle discrete time. There are other approaches that define continuous semantics for Dynamic Fault Trees in terms of Timed Markov Chains \cite{boudali2007compositional, junges2018one}. Our approach has a number of general advantages over these approaches: (i) we provide a rigorous definition of causality  and base our fault localisation on this definition, (ii) we directly employ data (in the form of trajectories) from the actual systems instead of building an intermediate representation, which, unless proven, may not be consistent with the actual system, (iii) our work is based on a generic semantic framework and can be instantiated for many different semantics for faults and conformance. We expect that the ideas  developed in this paper can be applied to those non-causal theories of fault localisation; however, this may require a reformulation of the basic concepts.


\section{Preliminaries}
\label{sec:preliminaries}

In this section, we first present the theory of actual causality in the context of discrete systems, originally proposed by Halpern and Pearl \cite{halpern2005causes}. Then, we provide a brief overview of cyber-physical systems and their models, present our running example and the motivation behind this work.

\subsection{Causal theory for discrete systems}
\label{sec:theoryDiscrete}

\begin{figure}[!h]
	\centering
	\includegraphics[width=1\textwidth]{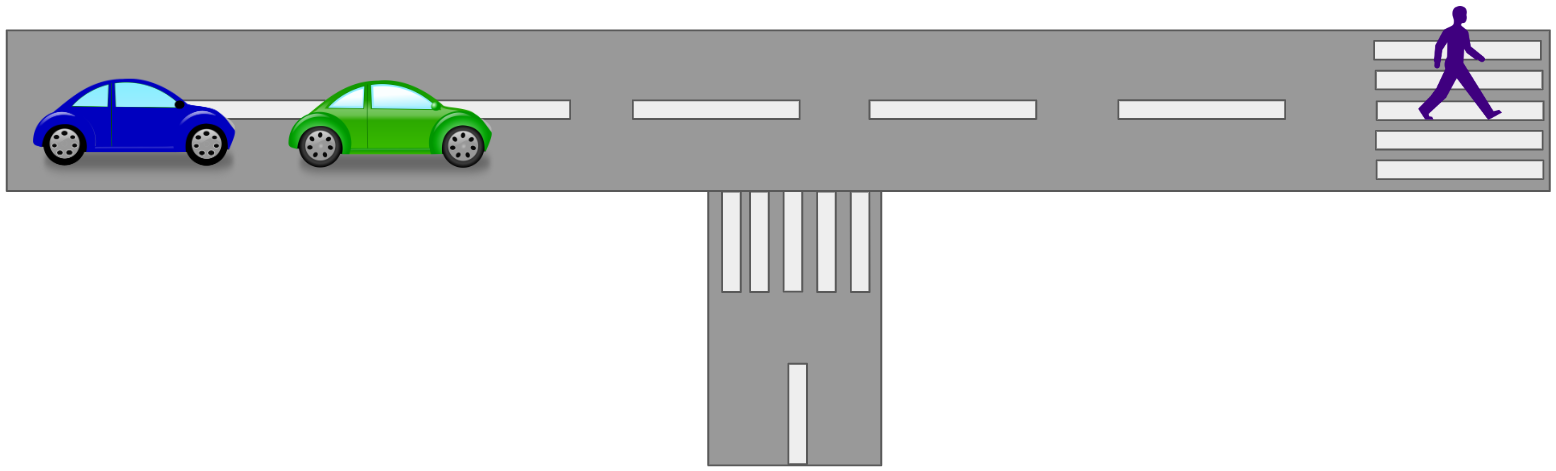}
	\caption{Illustration of the running example.}
	\label{fig:av-example}
\end{figure}

Consider the following discrete example (based on the classical Billy and Suzy example~\cite{halpern2016actual}): suppose that two autonomous vehicles A and B are driving on a straight road, one behind the other, in an extremely foggy weather condition. At the end of the road, a pedestrian is situated; the heavy fog prevents the vehicles' cameras from detecting the pedestrian. Furthermore, a junction on the road is situated before the pedestrian, which allows the vehicles to turn right. Here, we assume that the pedestrian cannot escape the imminent collision: if one of the vehicles does not turn right at the junction, then the pedestrian will be hit. In the case that neither vehicle turns right, vehicle A will hit the pedestrian but vehicle B will not. If vehicle A turns right, and vehicle B does not, then vehicle B will hit the pedestrian. If both vehicles turn right, the collision is avoided altogether. This scenario is depicted in Figure~\ref{fig:av-example}.

In this example, one can see that if neither of the two vehicles turn right, then only vehicle A can be the cause of the collision. Similarly, if vehicle A turns right, and B does not, then only vehicle B can be the cause.


Mathematical assessments of causality require formal modelling. As a precondition to a model, the signature provides the set of variables and their admissible valuations. Most of the formal definitions in this section are taken from those by Halpern and Pearl \cite{halpern2001causes}.

\begin{definition}[Signature]
	\label{def:signature}
	A signature is a tuple 
	\[\mathcal{S} = ( \mathcal{U},\mathcal{V},\mathcal{R} ), \]
    where $\mathcal{U}$ is a finite set of exogenous variables, $\mathcal{V}$ is a finite set of endogenous variables, and $\mathcal{R}$ associates with every variable $Y \in \mathcal{U} \cup \mathcal{V}$ a finite and non-empty set $\mathcal{R}(Y)$ of possible values for $Y$. 
\end{definition}

Exogenous variables are determined by factors outside of the model while endogenous variables are affected by exogenous ones and also by other endogenous variables. For instance, going back to the autonomous vehicle example, vehicle A and B turns can be seen as \textit{endogenous} variables, but the gravity and road friction that allow for the vehicles to steer can be seen as \textit{exogenous} variables.

\begin{definition}[Causal Model]
	\label{def:causalModel}
	A causal model over a signature $\mathcal{S}$ is a tuple 
	\[M = ( \mathcal{S}, \mathcal{F} ),\] where $\mathcal{F}$ associates with each variable $X \in \mathcal{V}$ a function denoted by $F_X$, such that:
	\[F_X: (\times_{U \in \mathcal{U}} \mathcal{R}(U)) \times (\times_{Y \in \mathcal{V} \backslash \{X\}} \mathcal{R}(Y)) \rightarrow \mathcal{R}(X)\]
\end{definition}

$F_X$ describes how the value of the endogenous variable $X$ is determined by the values of all other variables in $\mathcal{U} \cup \mathcal{V}$. The indexed cartesian products $\times_{U \in \mathcal{U}} \mathcal{R}(U)$ and $\times_{Y \in \mathcal{V} \backslash \{X\}} \mathcal{R}(Y)$ consider each possible values of the variables in $\mathcal{U}$ and $\mathcal{V} \backslash \{X\}$, respectively. Considering our discrete example, the causal model would have the following endogenous variables:

\begin{itemize}
    \item AT for (vehicle) A Turns: 1 if it turns right, and 0 if it does not.
    \item BT for (vehicle) B Turns: 1 if it turns right, and 0 if it does not.
    \item AH for A Hits: 1 if it hits the pedestrian, and 0 if it does not.
    \item BH for B Hits: 1 if it hits the pedestrian, and 0 if it does not.
    \item PH for Pedestrian Hit: 1 if the pedestrian is hit, and 0 if it is not.
\end{itemize}

The set $\mathcal{U}$ of exogenous variables comprises all information we need to assume so as to render all relationships deterministic (such as the presence of oxygen, gravity and the route the vehicles follow). We denote by $\vec{u}$  (i.e., a set of valuations in $\mathcal{R}(\mathcal{U})$) as the context of a cause. That is, the context is a mapping of exogenous variables to their values, which are used to induce the value of the endogenous variables. In our example, $\vec{u}$ can be seen as the context that makes the vehicle moving possible.

Furthermore, the types of causal models to which Halpern and Pearl restrict their definitions are called \emph{strongly recursive}. In essence, a causal model is strongly recursive, which for each endogenous variable, a context $\vec{u} \in \mathcal{R}(\mathcal{U})$ plays a role in defining its value. More specifically, the context helps defining the value of a subset of the endogenous variables, which, in turn, will be used in conjunction with the functions in $\mathcal{F}$ to determine the value of the remaining endogenous variables. 

In our discrete example, the context (which encompasses the route that each vehicle is following) comprises the variables $u_A \in \mathcal{U}$ and $u_B \in \mathcal{U}$. They represent the routes that vehicle A and B are following, respectively. They assume the value 1 if the respective vehicle is following a route that takes a right turn at the junction and 0 otherwise. In such a case, we can define the functions in $\mathcal{F}$ as follows.

\begin{itemize}
    \item $F_{AT}(\vec{u}, BT, AH, BH, PH) = u_A$ 
    \item $F_{BT}(\vec{u}, AT, AH, BH, PH) = u_B$
    \item $F_{AH}(\vec{u}, AT, BT, BH, PH) = \begin{cases} 
      0, & AT = 1 \\
      1, & AT = 0
   \end{cases}$
    \item $F_{BH}(\vec{u}, AT, BT, AH, PH) = 
    \begin{cases} 
      0, & AT = 0 \\
      1, & AT = 1 \wedge BT = 0
   \end{cases}$
    \item $F_{PH}(\vec{u}, AT, BT, AH, BH) = 
    \begin{cases} 
      0, & AH = 0 \land BH = 0 \\
      1, & AH = 1 \lor BH = 1
   \end{cases}$
\end{itemize}

In summary, the context considers the particular route that the cars are following, which dictates whether they will turn right or not (i.e., $AT$ and $BT$). Then, these variables affect $AH$ and $BH$, and those, in turn, affect $PH$, as defined in the functions ($\mathcal{F}$) above. 

\begin{figure}[!h]
	\centering
	\includegraphics[width=0.6\textwidth]{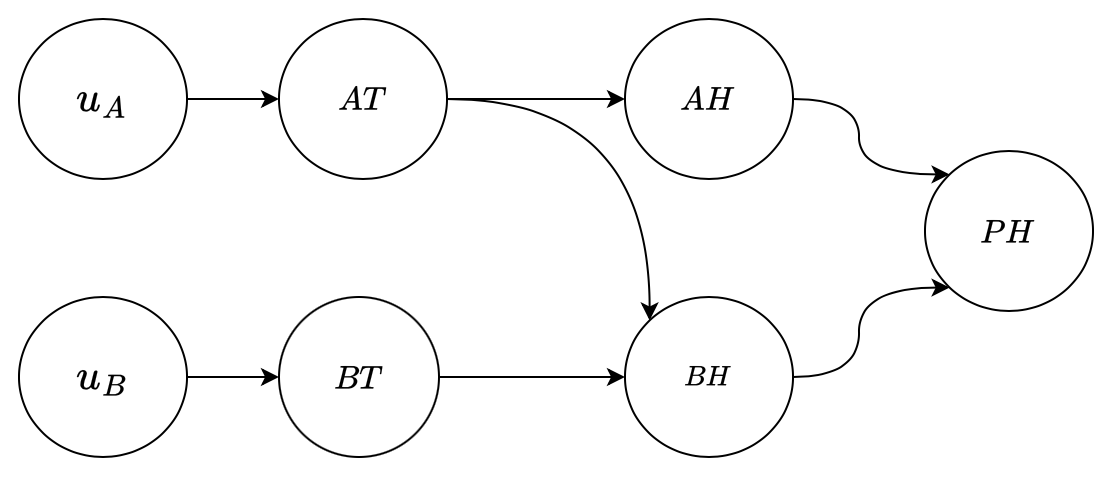}
	\caption{Causal graph of the AT/BT example.}
	\label{fig:graph-toy-example}
\end{figure}

In Figure~\ref{fig:graph-toy-example}, we display the causal graph of this example, in which the nodes (representing variables) that have a direct impact on each other are connected by an edge. 

The graphical representation of causal networks such as these are not used in the underlying theory nor in the implementation (see Section~\ref{sec:applications}); however, they provide a visual aid to understand the examples.

Finally, to make the definition of cause precise, we first need a syntax for causal events. Given a signature $\mathcal{S} = (\mathcal{U},\mathcal{V},\mathcal{R})$, a formula of the form $X = x$, for $X \in \mathcal{V}$ and $x \in \mathcal{R}(X)$, is called a primitive event.




\begin{definition}[Causal Formula]
	\label{def:basicCausalFormulaTrajectories}
	A causal formula is of the form
	\[ [X_1 \leftarrow x_1 , ... , X_k \leftarrow x_k]\Phi, where\]
	\begin{itemize}
	    \item $X_1,...,X_k$ are distinct variables in $\mathcal{V}$.
	    \item $x_i \in \mathcal{R}(X_i)$. And,
	    \item $\Phi$ is a Boolean combination of primitive events.
	\end{itemize}
\end{definition}

The formula $[X_1 \leftarrow x_1 , ... , X_k \leftarrow x_k]\Phi$ states that $\Phi$ holds in a system where $X_i$ is set to $x_i$ for $i = 1,...,k$. Such a formula can be abbreviated as $[\vec{X} \leftarrow \vec{x}]\Phi$.

\begin{definition}[Intervention]
	\label{def:causalModelUpdate}
    Given a causal model $M = (\mathcal{S},\mathcal{F} = \{\mathcal{F}_{X_1},\mathcal{F}_{X_2}...,\mathcal{F}_{X_k}\})$, and a set of assignments $[X_1 \leftarrow x_1 , ... , X_k \leftarrow x_k]$, an intervention on the causal model $M$, denoted by $M_{\vec{X} \leftarrow \vec{x}} = (\mathcal{S}, \mathcal{F}_{\vec{X} \leftarrow \vec{x}})$, is a modification to the structural equations in the causal model $M$ such that $\forall i \in \{1,...,k\}, \mathcal{F}_{X_i} = x_i$.

\end{definition}

An intervention of the type $X \leftarrow x$ can be interpreted as an update in $\mathcal{F}$ where the function for $X$ is set just to $x$. We define a satisfaction relation between causal model and causal formulae next.

\begin{definition}[Satisfaction Relation]
	\label{def:satisfactionRelationDiscrete}
    Given a causal model $M = (\mathcal{S} = (\mathcal{U} = \{U_1,...,U_m\},$ $\mathcal{V},\mathcal{R}),\mathcal{F})$, a context $\vec{u} = \{u_1,...,u_m\}$, and a primitive event $(Y = y)$, the satisfaction relation between the causal model, the context and the event, denoted by $(M,\vec{u}) \models (Y = y)$, holds if, and only if, $(\mathcal{F}_Y \in \mathcal{F} \land U_1 = u_1,..., U_m = u_m) \implies \mathcal{F}_Y = y $.

    Furthermore, given a causal formula $[\vec{X} \leftarrow \vec{x}](Y = y)$, the satisfaction relation between the causal model, the context and the causal formula, denoted by $(M,\vec{u}) \models [\vec{X} \leftarrow \vec{x}](Y = y)$, holds if, and only if, in the causal model resulting by the intervention $M_{\vec{X} \leftarrow \vec{x}} = (\mathcal{S}, \mathcal{F}_{\vec{X} \leftarrow \vec{x}})$, we have that $(\mathcal{F}_Y \in \mathcal{F}_{\vec{X} \leftarrow \vec{x}} \land U_1 = u_1, U_2 = u_2, ... , U_i = u_i) \implies \mathcal{F}_Y = y $.



\end{definition}

Thus, given a context $\vec{u} \in \mathcal{R}(\mathcal{U})$, we write $(M,\vec{u}) \models [\vec{X} \leftarrow \vec{x}](Y = y)$ if the variable $Y \in \mathcal{V}$ has the value $y$ in a causal model M where $X_i$ is set to $x_i$ for $i = 1,...,k$. The notation can also be used in the presence of a Boolean combination of primitive events: $(M,\vec{u}) \models [\vec{X} \leftarrow \vec{x}] \Phi$. Note that $(M,\vec{u}) \models [\vec{X} \leftarrow \vec{x}] \Phi \iff (M_{\vec{X} \leftarrow \vec{x}},\vec{u}) \models \Phi$.

Furthermore, in the special case where no assignments are performed (i.e., $k = 0$), we write $(M,\vec{u}) \models (Y = y)$, if and only if the variable $Y \in \mathcal{V}$ has the value $y$ given the context $\vec{u} \in \mathcal{R}(\mathcal{U})$ and the causal model $M$. This notation can also be used in the presence of a Boolean combination of primitive events: $(M,\vec{u}) \models \Phi$.


The types of events that are allowed as causes are of the form $(X_1 = x_1 \wedge ... \wedge X_k = x_k)$, that is, a conjunction of primitive events that can be abbreviated as $\vec{X} = \vec{x}$. Then, cause is formally defined as follows. 

\begin{definition}[Cause]
	\label{def:cause}
	We say that $\vec{X} = \vec{x}$ is a cause of $\Phi$ in $(M,\vec{u})$ if the following conditions hold:
	\begin{itemize}
	    \item AC1. $(M,\vec{u}) \models (\vec{X} = \vec{x}) \wedge \Phi$
	    \item AC2. There is a set $\vec{W}$ of variables in $\mathcal{V}$ and a setting $\vec{x}^\prime$ of the variables in $\vec{X}$ such that if $(M,\vec{u}) \models (\vec{W} = \vec{w})$, then:
        \[(M,\vec{u}) \models [\vec{X} \leftarrow \vec{x}^\prime, \vec{W} \leftarrow \vec{w}] \neg \Phi.\]
	    \item AC3. $\vec{X}$ is minimal; there is no strict subset $\vec{X'}$ of $\ \vec{X}$ such that $\vec{X'} = \vec{x'}$ satisfies conditions AC1 and AC2, where $\vec{x'}$ is the restriction of $\vec{x}$ to the variables in $\vec{X'}$.
	\end{itemize}
\end{definition}

In this definition, the variables in the set $\vec{W}$ allow for the cause to be tested under certain circumstances where the variables in $\vec{W}$ (which can be empty) are kept to their original values $\vec{w}$ even if they were supposed to be modified by the intervention $\vec{X} \leftarrow \vec{x}^\prime$. 

Consider the scenario in our autonomous vehicle example where $AT = 0, BT = 0, AH = 1, BH = 0,$ $PH = 1, u_A = 0,$ and $u_B = 0$. In our signature, we identify $AT, BT, AH, BH$ and $PH$ as endogenous variables ($\mathcal{V}$) and $u_A$ and $u_B$ as exogenous variables ($\mathcal{U}$); we would like to assess whether $AT = 0$ is the cause of the pedestrian being hit ($PH = 1$).

AC1 states that $(\vec{X} = \vec{x})$ cannot be a cause of $\Phi$, unless both the primitive causal events $(\vec{X} = \vec{x})$ and the effect $\Phi$ are true in the causal model M, given the context $\vec{u}$. That is, it states that for $AT = 0$ to be the cause of $PH = 1$, then both need to be true in $(M,\vec{u})$; thus, in this scenario, AC1 holds. Conversely, if we were trying to assess whether $BT = 1$ is the cause of $PH = 1$ then, AC1 could not be satisfied, as $BT = 1$ is not true in $(M,\vec{u})$ and, therefore, it could not be considered a cause.

AC2 states that for $(\vec{X} = \vec{x})$ to be a cause of $\Phi$, there must exist alternative values for the set $\vec{X}$ that leads to $\Phi$ not holding. This is under the assumption that there might exist a set of variables $\vec{W}$ that must be kept in its original value. To understand this concept, let us start with $\vec{W} = \{\}$ and assume we apply the intervention $AT \leftarrow 1$. This leads to vehicle A turning which results in vehicle B hitting the pedestrian ($BH = 1$) and, thus, the pedestrian is still hit ($PH = 1$); clearly, AC2 does not hold in this case. However, if we set $\vec{W} = \{BH\}$ and, hence, we maintain the value of variable $BH$ even under the intervention ($AT \leftarrow 1$) we have that neither car hits the vehicle ($AH,BH = 0$) and the pedestrian is not hit ($PH = 0$). Consequentially, we have that $(M,\vec{u}) \models [AT \leftarrow 1, BH \leftarrow 0] \neg(PH = 1)$ holds. This example shows the need for the set $W$.

Finally, AC3 asserts that the identified cause is minimal. In our scenario, it prevents ($AT = 0 \wedge BT = 0$) from being a cause, since ($AT = 0$) suffices to satisfy AC2. Thus, AC3 also holds and we can say that, in $(M,\vec{u})$, $(AT = 0)$ is a cause of $(PH = 1)$.



\subsection{Cyber-physical systems}
\label{sec:cps}

Cyber-Physical Systems (CPSs) integrate computational systems into their physical environments; examples of modern CPSs include vehicles and robotic systems \cite{mosterman2016cyber}. A typical CPS is a system where sensors feed input signals to a digital controller (discrete component) attached to physical actuators (continuous component) in a feedback loop. 

In order to model the continuous and discrete dynamics in CPSs, many formalisms have been used \cite{alur2015principles}. In this work, we make use of hybrid automata \cite{alur1993hybrid} to model the design of a CPS; it is a well-established formalism, with an intuitive semantics, besides being equipped with tools supporting different analyses \cite{annpureddy2011s,chen2013flow,frehse2011spaceex,fulton2015keymaera}. 

We first consider a running example. Then we introduce a motivation for our causal analysis of cyber-physical systems. Despite our choice of hybrid automata and Simulink in our mechanisation, our approach imposes no constraints on the formalism that our theory can be applied to.

\subsubsection{Running example: an autonomous vehicle}
\label{sec:runningExample}

An autonomous electric vehicle is driving at constant speed of 10 $m/s$ on a straight road, towards a stationary pedestrian situated on a crossroad. The braking distance ($d$) is a function that depends on the vehicle speed ($speed$), gravity ($g$) and the braking coefficient ($brakes$), which indicates the quality of the braking system (such as braking pads, tire quality, and tire pressure). Reasonable values for the latter are between 0.2 and 0.8.

\begin{equation*}
    d = \frac{speed^2}{2 * brakes * g}
\end{equation*}

Furthermore, the car is equipped with a lidar (a laser imaging, detection, and ranging system) that has 2 modes. A default long range mode detects objects within a 20 meters radius and a shorter range mode that halves the range. Whenever the battery enters a critical state, i.e., less than 5\% of the total charge, the lidar switches to short range mode to reduce power consumption.

Consider a scenario that, when t = 0, the car is 80 meters away from the crossroad where the pedestrian is stationed. The battery is at 10\% capacity and its consumption rate is constant at 1\% of the total capacity per second (i.e., it will decrease to $9\%$  after the 1 second, then to $8\%$ after 2 seconds, and so on). Table \ref{tab:av_inputs} describes the initial valuation for the system variables.

\begin{table}[!h]
\centering
\caption{Running example variables.}
\label{tab:av_inputs}
\begin{tabular}{|l|c|c|c|}
\hline
\rowcolor[HTML]{EFEFEF} 
\multicolumn{1}{|c|}{\cellcolor[HTML]{EFEFEF}Description} & \multicolumn{1}{|c|}{\cellcolor[HTML]{EFEFEF}Type} & \multicolumn{1}{|c|}{\cellcolor[HTML]{EFEFEF}Name} & \multicolumn{1}{c|}{\cellcolor[HTML]{EFEFEF} (Initial) Value} \\ \hline
Lidar range & Variable & \textit{lidarRange} & 20  \\ \hline
Critical battery threshold & Constant & \textit{critical} & 5 \% \\ \hline
Braking coefficient & Constant & \textit{brakes} & 0.2 \\ \hline
Car acceleration & Variable & \textit{acceleration} & 0 $m/s^2$ \\ \hline
Car speed & Variable & \textit{speed} & 10 m/s \\ \hline
Car position & Variable & \textit{carPosition} & 0 \\ \hline
Pedestrian position & Constant & \textit{pedestrianPosition} & 80  \\ \hline
Battery decay rate & Constant & \textit{decay} & 1\% / s \\ \hline
Battery charge left & Variable & \textit{battery} & 10\% \\ \hline
Gravity & Constant & \textit{g} & 9.8 $m/s^2$ \\ \hline
Critical check & Variable & \textit{belowCritical} & False \\ \hline

\end{tabular}
\end{table}


If one simulates the system, when $t \approx 8.5s$, the car collides with the pedestrian. We would like to know to which parts of the system this design flaw can be attributed; this challenge is akin to a verification problem. Regardless of whether the undesired behaviour is due to an actual system failure or an oversight in its specification, we aim to find the causes for it and correct the behaviour. 

We would like to note that the need for the distinction between constants and variables will be clear later on when interventions for cps are defined. In summary, a constant is a special variable that does not change value over time and, in this case, interventions can only change their value throughout the entire system execution.
\section{Causality for CPSs} 
\label{sec:theoryContinuous}

In this section, we first present the motivation behind this work and the formal definitions for the underlying theory that supports our strategy. The definitions presented in this section are based on the theory of actual causality discussed in Section \ref{sec:theoryDiscrete} but they are lifted to continuous systems.

\subsection{Analysis of cyber-physical systems}

Search-based test case generation techniques, such as the ones applied in our tool, HyConf \cite{araujo2019multiobjective}, can find faults in a system by searching for inputs that exercise extreme conditions on the System Under Test (SUT). However, the downside of such a methodology is the difficulty of tracing back the fault to a particular event or part of the system. 

It has been our experience that causes for faults in CPSs are very difficult to locate and, therefore, to fix. As a solution to this problem, causality is a concept with a potential to be exploited in the continuous setting. In the case of CPSs, our ultimate goal is to be able to determine which variables can be identified as causes of a fault and find the causes that exercise the safety levels of a system.

\begin{figure}[!h]
	\centering
	\includegraphics[width=0.35\textwidth]{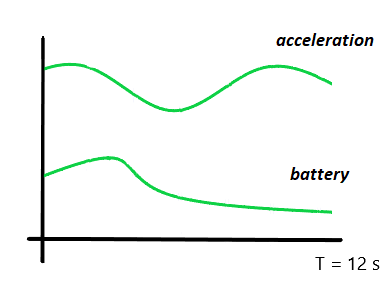}
	\caption{System trajectory.}
	\label{fig:causality_signals}
\end{figure}

In CPSs, one can express the system behaviour in terms of trajectories. A trajectory is the valuation of a set of variables over time. We first provide some informal intuition of trajectory and other notions; in Section 4.2 we formalise all these notions. Considering our running example, Figure \ref{fig:causality_signals} shows a trajectory of duration $T = 12s$ with hypothetical valuation for two of the system variables: the car acceleration and the battery charge.

\begin{figure}[!h]
\centering
\begin{subfigure}{.5\textwidth}
	\centering
	\includegraphics[width=0.8\textwidth]{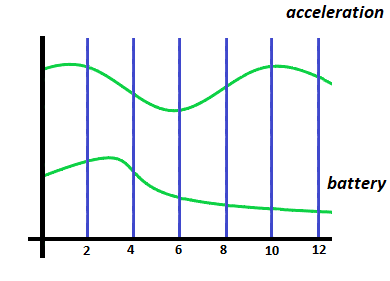}
	\caption{Trajectory split into time intervals  (slices).}
	\label{fig:causality_grid}
\end{subfigure}%
\begin{subfigure}{.5\textwidth}
  	\centering
	\includegraphics[width=0.8\textwidth]{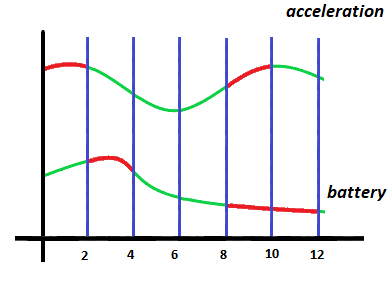}
	\caption{Highlighted cause and effect.}
	\label{fig:causality_causes}
\end{subfigure}
\caption{Trajectory and trajectory projection.}
\label{fig:trajectoryYandOverride_hypthetical}
\end{figure}

Our strategy for causal analysis is to split a trajectory of duration $T$ of a system into trajectory slices of equivalent size (see Figure \ref{fig:causality_grid}). We define trajectory slices as the projection of a trajectory considering a particular time interval. When building causal models, each trajectory slice is considered as a variable and, during the causal analysis, we use a set of slices to determine the cause of an events, such as the collision between the car and the pedestrian. For instance, in Figure \ref{fig:causality_causes}, the highlighted slices are the ones that could have been identified as the cause for the collision. 

When dealing with causal analysis for continuous systems, several aspects must be considered for a sound, yet realistic, theory and implementation. One of such aspects is the fact that determining cause is a costly computation~\cite{halpern2016actual}. Two important factors associated with the costs are (a) the number of endogenous variables in the causal model and (b) the number of variables in the cause. Thus, one cannot simply model a typical cyber-physical system by setting all variables as endogenous, otherwise determining the cause would be prohibitively computationally-intensive. In this work, in order to address factor (a), the choice of endogenous variables should be influenced by the variables that the user suspects are involved in the cause rather than the classical concept discussed in Section~\ref{sec:preliminaries}. Note that this does require some domain knowledge; however, selecting too many endogenous variables greatly affects performance (see our benchmarks in Section \ref{sec:benchmarks}). Moreover, in order to address factor (b), we consider an upper band to the number of variables in a cause and in the search; this is discussed in the mechanisation section (Section~\ref{ssec:process}).

On this topic, Beckers and Halpern~\cite{beckers2019abstracting} provide some insights into causal model abstraction which can allow modellers to think at a high level while still being faithful to a more detailed model. They discuss the notion of $\tau-$abstraction which suggests a transformation $\tau$ from causal model M1 to M2, where M2 is the high-level model where inessential differences are ignored.

Another topic to consider is with respect to cyclic models. In most intervention-based causal theories, the causal models must not comprise loops, such as in DAGs~\cite{geiger1990logic} and acyclic SCMs and GSEMs \cite{peters2021causal}. This has a significant impact on what can be modelled in CPS due to prominent presence of recursions and feedback loops. We circumvent this limitation by incorporating temporal dynamics into our causal models. This way, even in the presence of loops between system variables A and B (i.e., A affects the value of B and vice-versa), we use slices to build the causal models in such a way that a slice (i.e., a variable in a causal model) can only be affected by prior slices. 

In what follows, we define the supporting theory and present the mechanisation steps to achieve such goals.

\subsection{Trajectories and overriding} 

We start by defining valuation. Valuations serve as the basis for trajectories, which, in turn, define the semantic domain for models of CPSs.

\begin{definition}[Valuation]
Given a set of variables $\vec{V} = \{X_1, \ldots, X_n\}$, we denote by $\mathit{Val}(\vec{V}) = \vec{V} \rightarrow D$ the set of all total functions from $\vec{V}$ to the common domain $D$.
\end{definition} 

In the remainder of this paper, we take $D$ to be the set of real numbers $\mathbb{R}$. In cyber-physical systems, variables often have continuous valuation over time. This can be represented using trajectories, which are collections of variable valuations within a time interval. A discrete Boolean variable, for instance, can be modelled as a special case where the value taken by the variable is only 0 or 1.


\begin{definition}[Trajectory]
\label{def:trajectory}
Given a set of variables $\vec{V}$, the set of trajectories over $\vec{V}$,  denoted by $\mathit{Trajs}(\vec{V}) = \{x_1, \ldots, x_m\}$, is the set of all partial mappings $T \pfun Val(\vec{V})$, where $T$ is the time domain. 
\end{definition}

We take $T$ to be a convex subset of non-negative real numbers $\mathbb{R}_{+}$. We consider partial mappings over $T$, since trajectory slices (formally defined in the sequel), which may not be defined over the whole $T$, are also members of $\mathit{Trajs(\vec{V})}$. Below we define an auxiliary function that retrieves the set of variables in a trajectory, which is useful in later definitions. 

\begin{definition}[Variables of a Trajectory]
	\label{def:trajectoryVars}
We denote by $\mathit{var}(x)$, the set of variables $\vec{V}$ over which the trajectory $x$ operates.
\end{definition}

A trajectory that considers only a subset of the system variables can be obtained by projecting it over these variables.

\begin{definition}[Trajectory projection]
	\label{def:projection}
		Given a set of variables $\vec{V}$, the projection of a valuation $\mathit{val} \in \mathit{Val}(\vec{V})$ to $\vec{V}' \subset \vec{V}$, denoted by $\mathit{val} \downarrow_{\vec{V}'} \in \mathit{Val}(\vec{V}')$, is defined such that $\forall X \in \vec{V}'$, $(\mathit{val}\downarrow_{\vec{V}'})$ $(X) = \mathit{val}(X)$. Furthermore, the projection of a trajectory $x: T \pfun Val(\vec{V})$ to $\vec{V}' \subset \vec{V}$ is a trajectory $T \pfun Val(\vec{V}')$, denoted by $x \downarrow_{\vec{V}'}$,  such that $\forall t \in dom(x), (x \downarrow_{\vec{V}'})(t) = x(t) \downarrow_{\vec{V}'}$.
\end{definition}

\begin{example}[trajectory and projection]
\label{ex:trajectory}

Consider the running example discussed in Section \ref{sec:runningExample}. Figure \ref{fig:trajectory} shows a trajectory $x \in \mathit{Trajs}(\{battery,brakes,speed\})$. Figure \ref{fig:trajectoryProjection} shows the trajectory projection $x \downarrow_{\{battery\}}$.

\begin{figure}[!h]
\centering
\begin{subfigure}{.5\textwidth}
  \centering
	\includegraphics[width=0.8\textwidth]{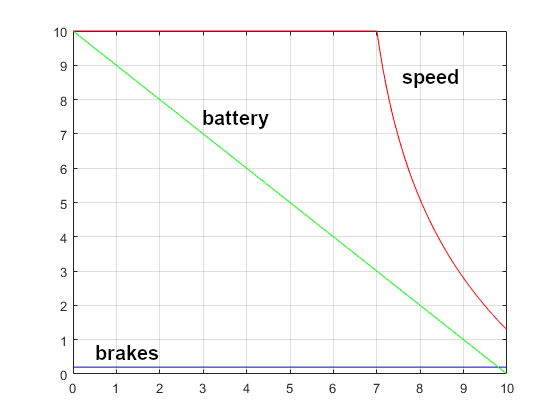}
	\caption{A trajectory \textit{x}.}
	\label{fig:trajectory}
\end{subfigure}%
\begin{subfigure}{.5\textwidth}
  \centering
	\includegraphics[width=0.8\textwidth]{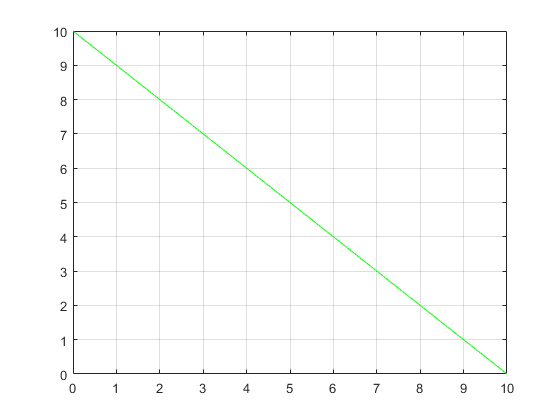}
	\caption{Trajectory projection $x \downarrow_{\{battery\}} $.}
	\label{fig:trajectoryProjection}
\end{subfigure}
\caption{Trajectory and trajectory projection.}
\label{fig:trajectoryYandOverride}
\end{figure}
\end{example}

Note that, since the braking coefficient is constant, the valuation of $brakes$ over time is a constant function. To discretise time in order to perform causal analysis and identify when the cause occurred, we first need to define the notion of time interval.

\begin{definition}[Time interval]
	\label{def:timeInterval}
	 A left-closed right-open interval $[i,j)$, where $i < j$, is defined as a convex subset of $\mathbb{R}_{+}$, such that, $\forall x \in \mathbb{R}_{+}, x \in [i,j) \iff i \leq x < j$.  
\end{definition}

In this work, a cause is defined using sub-trajectories of variables $X$ over one or more time intervals, as shown in Figure \ref{fig:trajectoryYandOverride_hypthetical}. To formalise that, we define trajectory slices.

\begin{definition}[Trajectory slice]
	\label{def:trajectorySlice}
	Given a set of variables $\vec{V}$, a trajectory $x \in \mathit{Trajs}(\vec{V})$, and an interval $[i, j)$ such that $[i, j) \subseteq \mathrm{dom}(x)$, a trajectory slice $x_{[i,j)} \in \mathit{Trajs}(\vec{V})$ is defined as a function $x_{[i,j)} : [i,j) \rightarrow Val(\vec{V})$, such that $\forall t \in [i,j), x_{[i,j)}(t) = x(t)$. 
	
\end{definition}

\begin{example}[trajectory slice]

Consider the trajectory $x$ from Example \ref{ex:trajectory}, Figure \ref{fig:trajectorySlice} shows the trajectory slice $x_{[4,6)}$ projected over the variable \textit{battery}, denoted as $x_{[4,6)} \downarrow_{\{battery\}}$.

\begin{figure}[!h]
	\centering
	\includegraphics[width=0.4\textwidth]{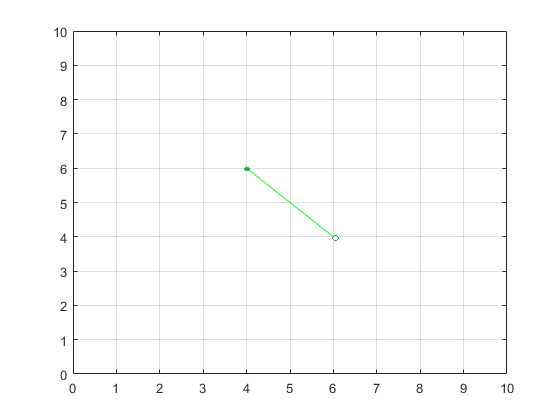}
	\caption{Trajectory slice projected over the variable \textit{battery}.}
	\label{fig:trajectorySlice}
\end{figure} 

\end{example}

\subsection{Causal models for CPSs}
\label{ssec:causal-models-CPSs}

As in the discrete case, a causal model is defined with respect to a signature.

\begin{definition}[Signature for cyber-physical systems]
	\label{def:signatureCPS}
	Given a set of variables $\vec{V}$, a signature for a cyber-physical system $C$, is a tuple 
	\[\mathcal{S}_{C} = \langle \mathcal{U},\mathcal{V},\mathcal{R} \rangle,\] \noindent where:
	\begin{itemize}
	    \item $\mathcal{U} \subseteq \vec{V}$ is a finite set of exogenous variables;
	    \item $\mathcal{V} \subseteq \vec{V}$ is a finite non-empty set of endogenous variables;
	    \item $\mathcal{R} \subseteq \mathit{Trajs}(\mathcal{U} \cup \mathcal{V})$.
	\end{itemize}
\end{definition}

In the discrete context, the set $\mathcal{R}$ contains the acceptable values for the variables in $\mathcal{U} \cup \mathcal{V}$. Analogously, in this extension, $\mathcal{R}$ is the set of selected trajectories that will be part of the causal analysis. In other words, we search for counterfactuals only within the set of trajectories $\mathcal{R}$. 

\begin{figure}[!h]
	\centering
	\includegraphics[width=0.9\textwidth]{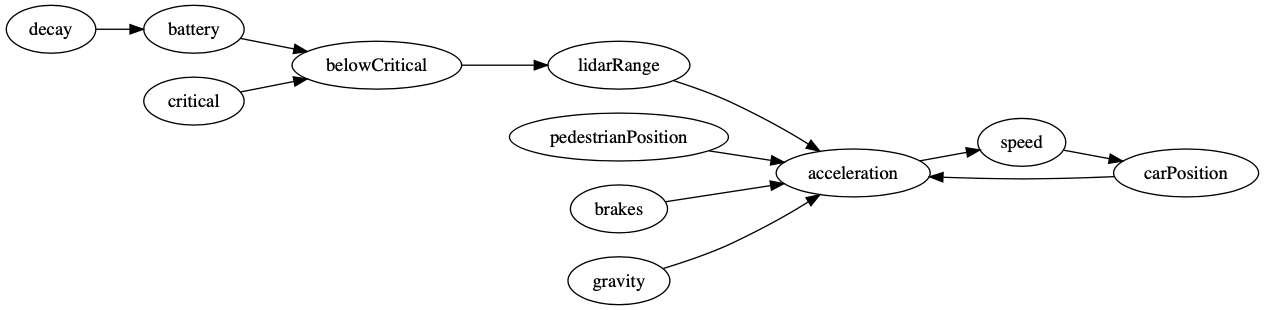}
	\caption{Cyclic causal graph of the running example.}
	\label{fig:graph-running-example-cyclic}
\end{figure}

\begin{figure}[!h]
	\centering
	\includegraphics[width=1\textwidth]{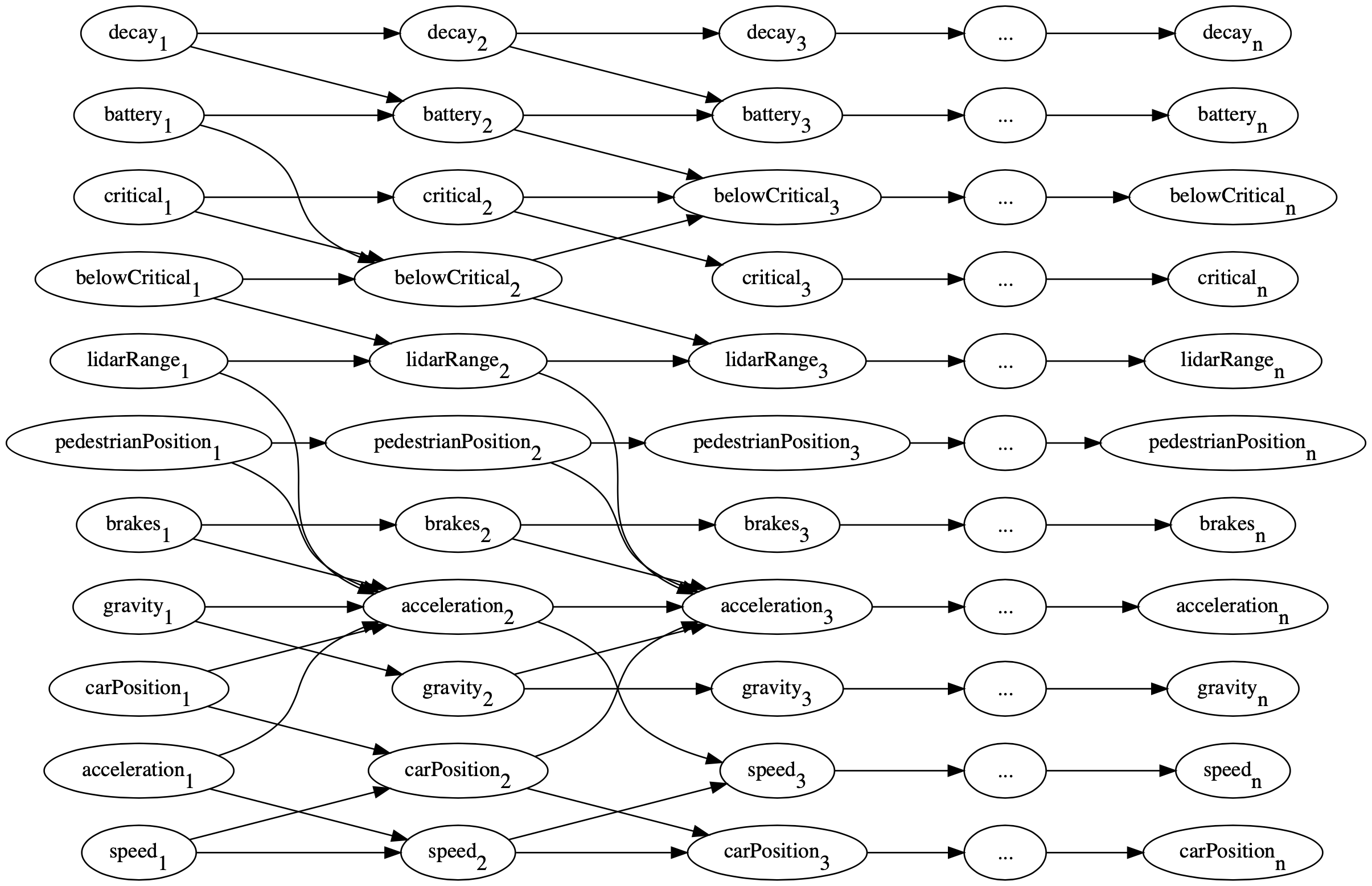}
	\caption{Acyclic causal graph of the running example.}
	\label{fig:graph-running-example-acyclic}
\end{figure}

Moreover, in our work, we assume that two properties always hold: the system must be deterministic and the causal model must be acyclic. We only consider deterministic systems since causal models are not achievable for non-deterministic systems; it would not be possible to define functions in $\mathcal{F}$ if variables could assume different values given the exact same parameter values. Furthermore, the dependency among endogenous variables in a causal model cannot be cyclic. That is, if the value of a variable $X$ affects a variable $Y$, then the opposite must not be true. Otherwise, when applying interventions to an endogenous variable $X$, that would effect changes in $Y$ and this, in turn, would result in further changes to $X$, which would conflict with the specific intervention that is being applied. This way, a cause could not be determined (as formalised in the sequel in Definition \ref{def:causeCPS}). This would result in some restrictions in systems that are inherently cyclic (e.g., with feedback loops) as only some specific combinations of endogenous variables are allowed. To mitigate this issue, we build causal models by splitting the system variables (e.g., $speed$) into slices and each slice corresponds to a causal variable (e.g., $speed_1$, $speed_2$,...,$speed_n$). Then, causes and effect are determined with respect to these slice variables. Figures~\ref{fig:graph-running-example-cyclic} and~\ref{fig:graph-running-example-acyclic} depict what would be causal model for the running example with and without using slices. Note how the cyyclic dependency between the variables $acceleration$, $speed$, and $carPosition$ is removed; with this causal mode, variables in the $n$th slice only affect the variables in the $n\text{th}+1$ slice. This results in the causal model being an abstraction of the real system, but for the purposes of verification, this suffices to determine actual causality given a sound implementation.


Moreover, most cyber-physical systems comprise a large number of variables and, thus, defining all these variables  as endogenous would render the analysis very costly in practice. Hence, in this work, the distinction between endogenous and exogenous variables depends on the feasibility and willingness to modify their values and assess causality. Hence, the best way to utilise this strategy require some knowledge about the system. Examples will be given throughout the paper to demonstrate this.


\begin{example}[Endogenous and exogenous variables]
\label{ex:end-ex-variables}

Consider the autonomous vehicle example given in Section \ref{sec:runningExample} and the hazard faced by the pedestrian and that the system is split into two slices. A possible choice of sets for endogenous and exogenous variables can be the following: 

\begin{itemize}
    \item $\mathcal{U} =$ \textit{\{g$_1$, g$_2$, critical$_1$, critical$_2$, lidarRange$_1$, lidarRange$_2$, carPosition$_1$, carPosition$_2$, pedestrianPosition$_1$, pedestrianPosition$_2$, decay$_1$, decay$_2$, acceleration$_1$, acceleration$_2$, belowCritical$_1$, belowCritical$_2$\}}
    \item $\mathcal{V} =$ \textit{\{battery$_1$, battery$_2$, speed$_1$, speed$_2$, brakes$_1$, brakes$_2$\}}
\end{itemize}

\end{example}

The causal model takes a signature and a set of functions; each function assigns a trajectory for an endogenous variable given a trajectory for the remaining system variables.

\begin{definition}[Causal model for cyber-physical systems]
\label{def:discretisedCausalModel}

Given a signature $\mathcal{S}_C = \langle \mathcal{U},\mathcal{V},\mathcal{R} \rangle$ for a cyber-physical system $C$, a causal model is defined as
 
\[M = \langle \mathcal{S}_C, \mathcal{F} = \{\mathcal{F}_X \mid X \in \mathcal{V} \}\rangle,\] 

\noindent where $\mathcal{F}_X: \mathit{Trajs}( \mathcal{U} \cup  \mathcal{V} \setminus \{X\}) \pfun \mathit{Trajs}(\{X\})$, such that $\mathrm{dom}(\mathcal{F}_X) = \{x \downarrow_{\mathcal{U} \cup \mathcal{V} \setminus \{X\}} \mid x \in \mathcal{R} \}$, and, for every trajectory $x \in \mathcal{R}$, $\mathcal{F}_X( x\downarrow_{\mathcal{U} \cup \mathcal{V} \setminus \{X\}}) = x \downarrow_{\{X\}}$.

\end{definition}

Intuitively, a causal model associates each variable $X \in \mathcal{V}$ with a single trajectory $x \downarrow_{\{X\}} \in \mathcal{R}$, given a trajectory for each of the other variables in $\mathcal{V}$. 

\begin{example} [A causal model for the running example]
\label{ex:deterministic}
Consider the example from Section \ref{sec:runningExample}, two trajectories $x$ and $y$ (Figure~\ref{fig:deterministic}), and a signature $\mathcal{S}_C = \langle \mathcal{U} = \{g, critical, lidarRange,$ $ carPosition, \ pedestrianPosition, \ decay, \ acceleration, \ belowCritical\}, \mathcal{V} = \{battery, speed,$ $brakes\},$  $\mathcal{R} = \{x,y\} \rangle$. Then, a causal model $M$ can comprise the following functions: 

\begin{itemize}
    \item $\mathcal{F}_{battery}(...) = 10 - (t*\textit{decay})$
    \item $\mathcal{F}_{brakes}(...) = 0.2$
    \item $\mathcal{F}_{speed}(...) = 10 + (t * acceleration)$
\end{itemize}

\begin{figure}[!ht]
  \centering
  \begin{subfigure}[b]{0.4\linewidth}
    \centering
    \includegraphics[width=1\linewidth]{Figures/bad_breaks_bad_battery.png} 
    \caption{Trajectory $x$.} 
    \label{fig:deterministic1} 
  \end{subfigure}
  \begin{subfigure}[b]{0.4\linewidth}
    \centering
    \includegraphics[width=1\linewidth]{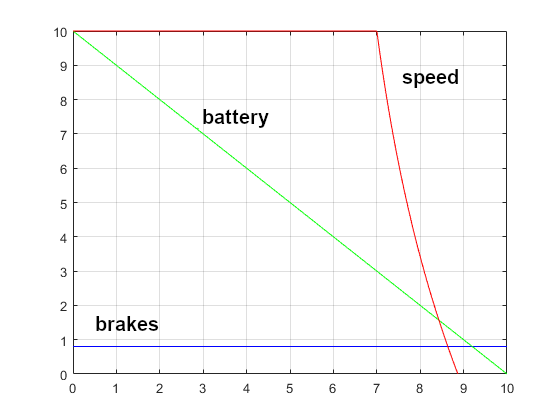} 
    \caption{Trajectory $y$.} 
    \label{fig:deterministic12} 
  \end{subfigure}
  \caption{Trajectories of the running example.}
  \label{fig:deterministic}
\end{figure}

\end{example}

For simplification, we omit the variable names in the parameter field of the functions in the example above. In the next section, the formal syntax to represent causes and effects are given as Boolean expressions between trajectories. Therefore, we first define a grammar for a primitive event and then we describe its semantics informally.


\vspace{\baselineskip}

\noindent \textbf{Primitive Event Grammar}

\begin{grammar}

<Expression> ::= <Trajectory> 
                 \alt <Trajectory> + <Expression>
                 \alt <Trajectory> - <Expression>
                 \alt <Trajectory> * <Expression> 
                 \alt <Trajectory> / <Expression> 

<BooleanOperator> ::= $=_{<Interval>}$
               \alt $\textless_{<Interval>}$
               \alt $\textgreater_{<Interval>}$
               \alt $\geq_{<Interval>}$
               \alt $\leq_{<Interval>}$

<PrimitiveEvent> ::= <Trajectory> <BooleanOperator> <Expression>

<PrimitiveEventExpression> ::= <PrimitiveEvent> 
                               \alt $\neg$ <PrimitiveEvent>
                               \alt <PrimitiveEvent> $\land$ <PrimitiveEventExpression>
                               \alt <PrimitiveEvent> $\lor$ <PrimitiveEventExpression>

\end{grammar}

In our logic, a trajectory is assigned to a variable (e.g., $x : T \rightarrow Val(battery)$) but can also be represented by constant (e.g., $x : T \rightarrow 3.14$). An expression, however, is defined as arithmetic operations between trajectories. Given two trajectories $x,y$ result between the addition $x + y$ is a new trajectory $s | \forall t \in dom(x) \cap dom(y) : s(t) = x(t) = y(t)$. The arithmetic operations between trajectories are defined over the addition, subtraction, multiplication and division operators.

Lastly, a primitive event is a Boolean comparison between a trajectory and an expression.
Given two variables $X$ and $Y$, two trajectories $x \in \mathit{Trajs}(\{X\})$ and $y \in \mathit{Trajs}(\{Y\})$, and an interval $[i,j) \subseteq dom(x) \cap dom(y)$, the primitive event of the type $x =_{[i,j)} y$ holds when for all $t \in [i,j)$, we have that  $x(t) = y(t)$.


 For instance, we use $x =_{[i,j)} y + z$ as syntactic sugar for $x =_{[i,j)} s \mid \forall t \in [i,j) : s(t) = y(t) + z(t)$, and, hence, we have that $\forall t \in [i,j) : x(t) = y(t) + z(t)$. Similarly, we also use real valued numbers in a primitive event (which can be seen as a new trajectory that is constant and set at that value). As an example, the primitive event $x <_{[i,j)} y * 3.14$ is syntactic sugar for $x <_{[i,j)} s \mid \forall t \in [i,j) : s(t) = y(t) * 3.14$.

In this extension, we denote $\Phi$ (i.e., the effect) as a Boolean combination of primitive events for CPSs. Given a causal model $M = \langle \langle \mathcal{U}, \mathcal{V}, \mathcal{R} \rangle, \mathcal{F} \rangle$, a trajectory $c : \mathit{Trajs}(\mathcal{U} \cup \mathcal{V})$ and a set of trajectory slices $\vec{x} = \{x \mid x \in \mathit{Trajs}(\mathcal{V})\}$, causes for CPSs are given in the format $(c \mapsfrom \vec{x})$. The special case where a conjunction of primitive events using the equality operator between one trajectory $(c)$ and each trajectory slice in a set ($\vec{x} = \{x_1,x_2,\dots,x_n\}$) is denoted by $(c \mapsfrom \vec{x})$. This format is an abbreviation for $\bigwedge_{i=1}^n (c \downarrow_{var(x_i)} =_{\mathit{dom}(x_i)} x_i)$. In practice, the trajectory $c$ is obtained from a system execution and the trajectory slices $\vec{x}$ are taken from $c$. Hence, a cause $(c \mapsfrom \vec{x})$ means that the fact that the output of the system $c$ comprises the specific trajectory slices $\vec{x}$ is a cause to an effect $\Phi$.

In the discrete context, in order to identify causes, one needs to apply interventions and modify the value of variables. In the context of CPSs, causes are given using trajectory slices and thus the interventions need to be applied on the same slices as the prospective cause. Hence, we formally define the concept of alternative set of trajectories. Given a set of trajectories $(\vec{x})$, an alternative set of trajectories $(\vec{x}^\prime)$ is one that, for each trajectory in $\vec{x}$, there exists a corresponding trajectory in $\vec{x}^\prime$ that ranges over the same variable(s) and the same time interval but differ in value. The opposite must also be true.

\begin{definition}[Alternative set of  Trajectories]
	\label{def:alternativeTrajs}
	
	Two sets of trajectories $\vec{x}$ and $\vec{x}^\prime$ are called alternative if, and only if: 
	  
    \begin{itemize}
    \item $\forall x \in \vec{x} : \exists! x^\prime \in \vec{x}^\prime : (var(x) = var(x^\prime) \wedge dom(x) = dom(x^\prime) \wedge \neg(x^\prime =_{dom(x)} x))$

    \item $\forall x^\prime \in \vec{x}^\prime : \exists! x \in \vec{x} : (var(x) = var(x^\prime) \wedge dom(x) = dom(x^\prime) \wedge \neg(x^\prime =_{dom(x)} x))$
        
    \end{itemize}

\end{definition}

We denote by $\mathit{Alts}(\vec{x})$ the set of all alternative set of trajectories of $\vec{x}$. Lastly, we lift the definition of the satisfaction relation (Definition~\ref{def:satisfactionRelationDiscrete}) to work with causal models for cyber-physical systems. Similarly to the discrete version, we use $(M,u)$ to represent a causal model and a context, which is now represented by a trajectory $u : \mathit{Trajs}(\mathcal{U})$ over the exogenous variables.

\begin{definition}[Satisfaction relation for cyber-physical systems]
	\label{def:satisfactionRelation}
    Given a causal model for cyber-physical systems $M = \langle \mathcal{S_C} = \langle\mathcal{U} = \{U_1,U_2,...,U_m\},$ $\mathcal{V},\mathcal{R}\rangle,\mathcal{F}\rangle$, a trajectory $u \in \mathit{Trajs}(\mathcal{U})$, and a primitive event $x =_{[i,j)} y$ where $x \in \mathit{Trajs}(\{X\})$ and $y \in \mathit{Trajs}(\{Y\})$, and $X,Y \in \mathcal{U} \cup \mathcal{V}$, the satisfaction relation between the causal model, the context, and the event, denoted by $(M,u) \models (x =_{[i,j)} y)$, holds if, for all $t \in [i,j)$, we have $(U_1 = u\downarrow_{U_1}, U_2 = u\downarrow_{U_2}, ... , U_m = u\downarrow_{U_m}) \implies (x(t) = y(t))$.
\end{definition}

Trivially, the satisfaction relation can be extended to a Boolean combination of primitive events, such as causes $(c \mapsfrom \vec{x})$ or effects ($\Phi$).

\subsection{Causes for CPSs} 

We define below the notion of causal model update, which is akin to an intervention in the discrete case.

\begin{definition}[Causal model update]
	\label{def:causalModelUpdate}
	Given a signature $S_C = \langle \mathcal{U}, \mathcal{V}, \mathcal{R}\rangle$,  a set of variables $\vec{X} \subseteq \mathcal{V}$ and a set of trajectories $\vec{x} = \{x \mid x \in \mathit{Trajs}(\vec{X})\}$, an updated signature, denoted by $S_C^{'}$, is defined as $\langle \mathcal{U}, \mathcal{V}, \mathcal{R}^{\vec{X} \leftarrow \vec{x}}\rangle$, where $\mathcal{R}^{\vec{X} \leftarrow \vec{x}}$ is the set of all trajectories $r \in \mathcal{R}$ such that $r \downarrow_{\{X\}} =_{\mathit{dom}(x)} x \downarrow_{\{X\}}$, for each $x \in \vec{x}$.
	
	Given a causal model $M_C = \langle S_C, \{\mathcal{F}_Y \mid Y \in \mathcal{V}\} \rangle$, an updated causal model, denoted by $M_{\vec{X} \leftarrow \vec{x}}$, is defined as $\langle S_C^{'}, \{\mathcal{F}_Y^{'} \mid Y \in \mathcal{V}\} \rangle$, where the updated function for $\mathcal{R}^{\vec{X} \leftarrow \vec{x}}$, denoted as $\mathcal{F}_Y^{'}$, is defined on $\{x \downarrow_{U \cup V \setminus \{X\}} \mid x \in \mathcal{R}^{\vec{X} \leftarrow \vec{x}} \}$.
\end{definition}

Intuitively, $M_{\vec{X} \leftarrow \vec{x}}$ is an updated causal model obtained by filtering the trajectories of the variables in $\vec{X}$ with the trajectories in $\vec{x}$. A sequence of two causal model updates is denoted by $M_{\vec{X} \leftarrow \vec{x},\vec{Y} \leftarrow \vec{y}}$.

To determine cause, similarly to the discrete version, we use $(M,u)$ to represent a causal model and a context, which is represented by a trajectory $u \in \mathit{Trajs}(\mathcal{U})$ over the exogenous variables. Now, consider a trajectory $c$ that led to $\Phi$ where $c \downarrow_{\mathcal{U}} =_{\mathrm{dom}(u)} u$, then, a cause of $\Phi$ in $(M,u)$, given in the format $(c \mapsfrom \vec{x})$, is determined as follows.

\begin{definition}[Cause for cyber-physical systems]
	\label{def:causeCPS} 
	Given a causal model $M = \langle \langle \mathcal{U}, \mathcal{V}, \mathcal{R} \rangle, \mathcal{F} \rangle$, a setting $u \in \mathit{Trajs}(\mathcal{U})$, a trajectory $c : \mathit{Trajs}(\mathcal{U} \cup \mathcal{V})$ such that $c\downarrow_{\mathcal{U}} =_{\mathrm{dom}(u)} u$, and a set of trajectory slices $\vec{x} = \{x \mid x \in \mathit{Trajs}(\mathcal{V})\}$, then $(c \mapsfrom \vec{x})$ is a cause of $\Phi$ in $(M,u)$ when the following three conditions hold:
 
	\begin{itemize}
	    \item AC1. $(M,u) \models (c \mapsfrom \vec{x}) \wedge \Phi$,
	    
	    \item AC2. There exists a set of variables $\vec{W} \subset \mathcal{V}$ and two sets of trajectories $\vec{x}^\prime \in \mathit{Alts}(\vec{x})$, and $\vec{w} \subseteq \mathit{Trajs}(\vec{W}) \cap \mathcal{R}$ such that if $(M,u) \models (c \mapsfrom \vec{w})$, then:

        \[(M_{\vec{X} \leftarrow \vec{x}^\prime, \vec{W} \leftarrow \vec{w}}, u) \models \neg \Phi\]

	    \item AC3. There is no strict subset of $\vec{x}$ that satisfies AC1 and AC2.
	\end{itemize}
\end{definition}

This extension of the original definition for discrete systems (see Definition \ref{def:cause}) fully considers continuous trajectories and time intervals, with cause being established using trajectory slices. We aimed for a conservative extension, however a few things are to be considered.

Firstly, in the original definition, Halpern uses $\vec{u}$ to describe the set of exogenous variables and valuations. In our definition, a trajectory contains multiple variables and their valuation over time, thus a trajectory $u \in \mathit{Trajs}(\mathcal{U})$ that contains all exogenous variables and their valuation over time is used.

Secondly, the minimality clause, AC3, now focus on trajectory slices rather than variables. In the original theory, AC3 states that no subset of the \emph{variables and their specific values} in the prospective cause should satisfy AC2. In our extension, however, we employ trajectory slices, which are already defined based on both variables and values. Thus, given that a particular list of trajectory slices ($\vec{x}$) satisfies both AC1 and AC2, then no subset of $\vec{x}$ should satisfy AC1 and AC2 in order to satisfy AC3.

Thirdly, our theory does not restrict $\Phi$ to only endogenous variables. In the original work, strictly speaking, endogenous variables could not affect exogenous ones, thus they could not be included in $\Phi$. In this work, however, this is not needed nor included in the definition as the choice of endogenous variables is a user choice in order to make the performance of the causality checks viable in practical settings (more on that in Section~\ref{sec:benchmarks}).

Lastly, we note how the set $\vec{w}$ is a subset of $\mathit{Trajs}(\vec{W}) \cap \mathcal{R}$. While this may seem counter-intuitive, the reason is that this enables a sound mechanisation of this theory. Suppose one is to determine whether $(c \mapsfrom \vec{x})$ is a cause of a $\Phi$. This requires one to evaluate whether, for all possible variables in $\vec{W}$, there exists any set of trajectory slices $\vec{w}$ that would result in a subset of $\vec{x}$ satisfying AC1 and AC2. This is intractable. Hence, instead, a mechanisation of this theory can restrict the set $\mathcal{R}$ to a countable set of trajectories (found, for instance, by a search heuristic).


We use Example~\ref{ex:deterministic} to explain Definition~\ref{def:causeCPS}. Consider that the trajectory $x$ from Figure~\ref{fig:deterministic1} is $c$, that is, the scenario where a failure was observed. We define $\Phi$ as $\neg (c \downarrow_{\{speed\}}$ $\leq_{[8.9,9)} 0)$, which holds if the variable $speed$ is greater than $0$ at any point in the interval $[8.9,9)$. Essentially, we are checking the causes for the vehicle not stopping around the 9 seconds mark.

Now, we determine whether $c \mapsfrom \{ x \downarrow_{\{brakes\}} \}$ is a cause for $\Phi$. Clearly, AC1 holds; both the cause and the effect are true in the model. Further, the cause is a singleton, which means it is minimal; thus AC3 holds. As for AC2, the set $\vec{W}$ can be empty. Consider that $\vec{X} = \{brakes\}$ and $\vec{W} = \{\}$ and consider the trajectory $\{y \downarrow_{\{brakes\}}\}$ (from Figure~\ref{fig:deterministic12}) as our alternative set of trajectories. It is clear that AC2 is satisfied as $(M_{\vec{X} \leftarrow \vec{y}} \downarrow_{\{brakes\}},u) \models \neg \Phi$ holds. We can see in Figure~\ref{fig:deterministic12} that, when we increase the brakes, the car fully stops before the 9 seconds mark. We note that constants do not change over time and, thus, the value for the braking coefficient is set at the beginning of the scenario and is the same throughout. When an intervention is applied to such a variable, its value must change for the entire duration of its trajectory.

We further exemplify our theory of causality for CPSs (particularly, Definition 16) in Section 5.2. We provide our algorithm for causal assessment and use another concrete example for its explanation.
\section{Practical applications of causality}
\label{sec:applications}

In this section, we demonstrate how causal analysis can be applied in practice to CPSs. First, we detail a standalone process for causal analysis and then we exemplify, using the running example, its applicability in the verification process.

\subsection{Causal analysis process}
\label{ssec:process}

We explain the step-by-step process for causal analysis. The process incorporates the theory defined in Section~\ref{sec:theoryContinuous} and has been mechanised and integrated into HyConf\footnote{https://github.com/hlsa/HyConf}, a tool for test case generation and conformance testing of cyber-physical systems. HyConf employs multi-objective search-based heuristics to find challenging input trajectories that exercise extreme conditions of the system under test (SUT). Particularly, it checks for conformance violations based on the $(\tau,\epsilon)$-conformance notion \cite{abbas2014conf}. In short, given two output trajectories (e.g., the system implementation and an idealised specification model), the tools compares the distance between the two trajectories in terms of time ($\tau$) and space ($\epsilon$). These parameters can be seen as margins of error, and if the two trajectories are beyond such values, the a non-conforming verdict is given (i.e., a failure was observed). The inputs and failures observed in future sections were obtained via HyConf. These are, in turn, fed as input to our causal analysis.

\begin{figure}[!t]
    \centering
    \includegraphics[width=1\textwidth]{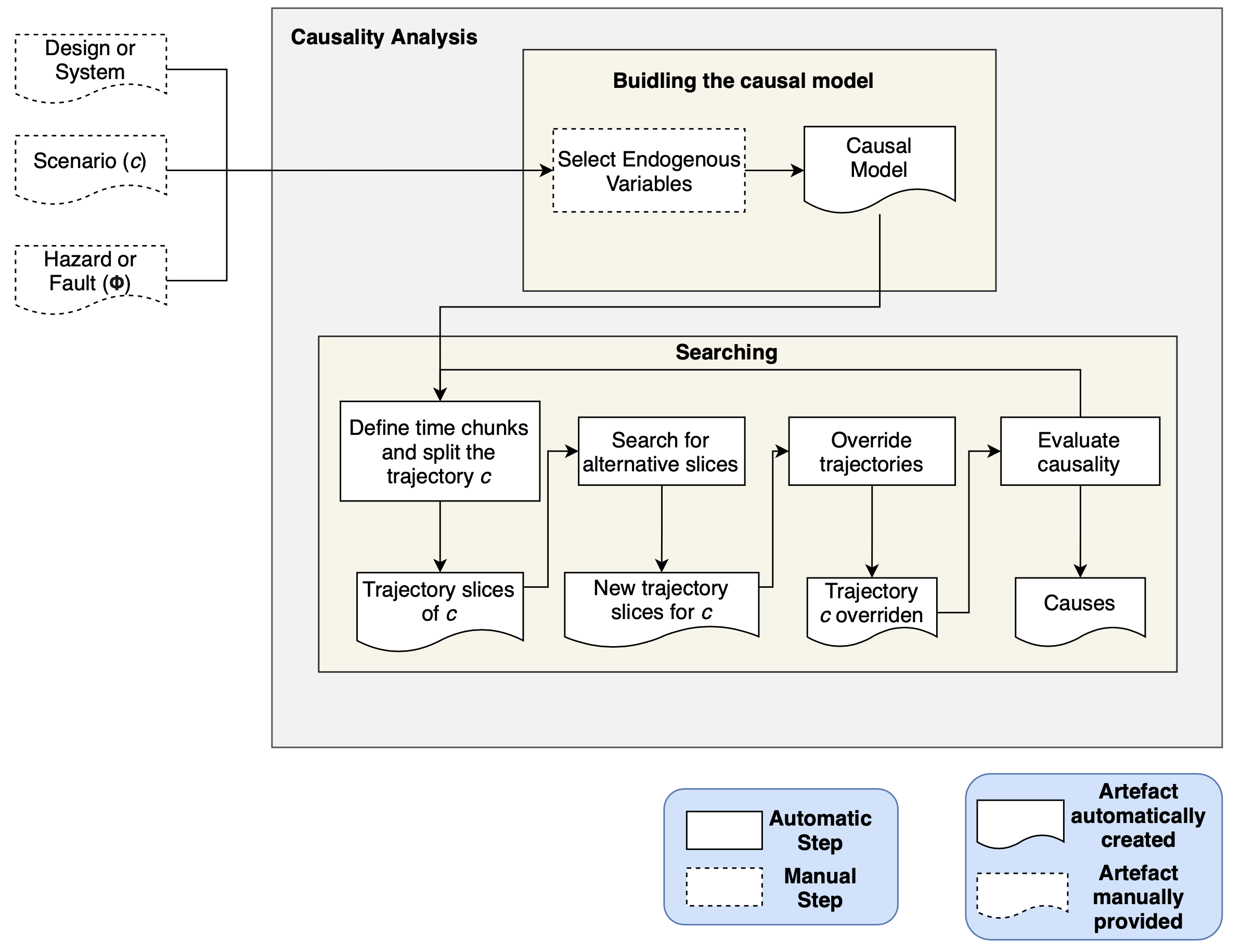}
    \caption{Causality process.}
    \label{fig:causalityProcess}
\end{figure}

We consider the running example (see Section \ref{sec:runningExample}) and use it to illustrate our approach. Figure \ref{fig:causalityProcess} depicts an overview of the process. The dotted lines represent manual steps or manually provided (input) artefacts whilst solid ones represent automatic steps or automatically provided artefacts, as explained in the bottom right of the figure. Our causal analysis process requires 3 inputs from the user: the system (implementation) or design (model) under verification (currently, we accept Matlab/Simulink models), the hazard or fault ($\Phi$, written in a simplified format of Signal Temporal Logic formulae \cite{maler2004monitoring}) and the scenario (a trajectory $c$, in Matlab/Simulink output format) in which the hazard/fault occurs. What we call a scenario in this work is a finite execution of a system in the form of a trajectory that comprises all variables within the system. From this scenario, we derive the valuation of both the endogenous and the exogenous sets (including the context setting $u$) of variables.

The step-by-step is given below.

\begin{enumerate}
    \item A model of a cyber-physical system, a scenario and the hazard must be given.
    \item The user chooses the set of endogenous variables and the remaining variables are assumed exogenous automatically.
    \item The causal model is built with a singleton $\mathcal{R}$ that (initially) only contains the trajectory $c$ that led to the hazard.
    \item The search commences by automatically splitting the trajectory $c$ (2 time intervals, initially) and therefore dividing it into smaller (and still continuous) trajectory slices.
    \item The search finds new trajectory slices and then these are used to override $c$ to look for violations of $\Phi$. $\mathcal{R}$ is updated on-the-fly.
    \item Causality assessment is performed. If a cause is found, the time intervals are reduced to increase precision. 
    \item If a maximum number of trajectory intervals has been explored, then the process stops. Otherwise, the search continues with a finer granularity of time intervals.
\end{enumerate}

The first stage is to build a causal model. It is important to emphasise that, in a practical setting, our causal models are an approximation of the complex interactions of the physical system. Thus, there are some simplifications in the way that the causal models are built.

\begin{itemize}
    \item In our theory, every variable that is influenced by the system should be classified as endogenous. However, in our practical causal models, having too many endogenous variables would render the analysis very lengthy and costly. Thus, we leave the choice of endogenous variables to the user's discretion and to what they may want to analyse as cause.
    
    \item In our implementation, the set $\mathcal{R}$ is built iteratively. We start with a singleton $\mathcal{R}$ that only contains the trajectory that led to the fault ($c$); as the search evaluates new trajectories, by overriding $c$ with trajectory slices in order to find a cause, they are added to $\mathcal{R}$. Once the search is concluded, multiple causes may have been found and, thus, we assess them considering the final $\mathcal{R}$ that has been built; this is the moment where we apply our theory for causality assessment of CPSs (see Section~\ref{sec:theoryContinuous}).
    
    \item When users choose the endogenous variables, they must set a maximum and minimum value and, if appropriate, mark the variable as constant. This is to avoid finding causes that do not respect system requirements and dynamics.
    
\end{itemize}

We re-iterate that the equations and variables of the causal model are provided as inputs from the user and the set $\mathcal{R}$ (state space) is built during the search. Moreover, causes are determined with respect to a particular causal model. Setting the maximum and minimum values for variables binds the set $\mathcal{R}$ of the causal model, and therefore the results are sound with respect to that particular model. Due to manual inputs (such as the boundaries for system variables), it is possible for the causal model to not reflect the actual system (from which $c$ is taken) and, in this case, the cause might not reflect the reality of it. However, as far as cause/causal model are concerned, the results are still sound. In this work, we assume that the causal model that is built (given the manual inputs) respects the reality of the system, that is, one has to make sure that the boundaries in the causal model respect the boundaries in the physical system.

Furthermore, unlike variables, constants cannot be overridden in time intervals, only in their whole duration. This can be seen as a restriction on the set $\mathcal{R}$, such that constants can only be defined by constant trajectories. The set of functions $\mathcal{F}$ that describe how the endogenous variables are affected by other system variables must be provided by the user. Given this information, the tool checks whether the causal model is acyclic, otherwise the causality assessment becomes infeasible, as explained at the end of Section~\ref{ssec:causal-models-CPSs}.

A difficulty associated with this strategy is the number of combinations of trajectory slices that have to be analysed to find the causes. This is particularly influenced by the chosen granularity of time intervals. That is, the number of time intervals increases the time spent in the search but conversely can identify a cause with more precision.

Furthermore, since the causes of a failure are more likely to be closer to the moment in time when the failure happened, an attenuation factor is implemented, where the search assigns higher search priority to the slices closer to the fault. In what follows, we present two algorithms employed in our strategy and we explain them using an example. Algorithm~\ref{alg:causalityAlgorithm} shows the pseudo-code that conducts the causal analysis and Algorithm~\ref{alg:causalitySearchAlgorithm} shows the psuedo-code for the search that employs the causal analysis algorithm. 

\subsection{Application to the running example}
\label{ssec:verification}

The search algorithm (shown in Algorithm~\ref{alg:causalitySearchAlgorithm}) receives the causal model ($M_C$), the hazard ($\Phi$) and the trajectory $c$ that leads to $\Phi$, as input; its output is a set of causes. In what follows, we apply the algorithm to our running example (the autonomous vehicle) to find causes for the collision hazard. First, we define the inputs to the algorithm.

\subsubsection{Defining the hazard} The given cyber-physical system $C$ is modelled in Simulink based on the example presented in Section \ref{sec:runningExample}. We characterise the hazard as the collision between the car and the pedestrian. The pedestrian position ($pedestrianPosition$) is set to 80 whilst the initial position of the car ($carPosition$) is set to 0. The collision occurs if the car cannot brake in time and its position exceeds that of the pedestrian (i.e., $carPosition \geq pedestrianPosition$).

\begin{table}[!h]
    \centering
    \caption{Issue observed}
	\label{tab:runningFaults}
	\begin{tabular}{|c|c|}
		\hline
		\cellcolor{Gray} \textbf{Effect} & \cellcolor{Gray} \textbf{$\Phi$} \\ \hline
		The car collides with the pedestrian & $c \downarrow_{\{carPosition\}} \geq_{\mathrm{dom}(c)} c \downarrow_{\{pedestrianPosition\}}$ \\ \hline
	\end{tabular}
\end{table}

Table \ref{tab:runningFaults} describes the effect with respect to which the cause is being established, along with $\Phi$. Currently, the tool only accepts expressions using logical and arithmetic operators, logical connectors ($\wedge$, $\vee$), real numbers, and the variable names.

\begin{figure}[!h]
\centering
\includegraphics[width=0.5\linewidth]{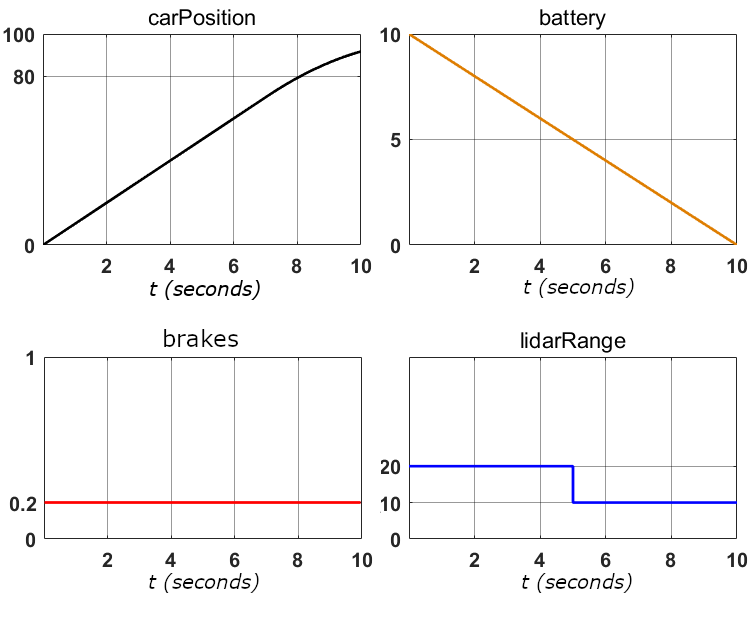} 
\caption{Projections of trajectory $c$ that leads to $\Phi$.} 
\label{fig:originalTrajectory} 
\end{figure}

If we execute the system as originally defined in Section \ref{sec:runningExample}, the battery enters the critical state and the lidar range is reduced. Then, the car detects the pedestrian but does not brake in time and a collision occurs (i.e., $carPosition \geq 80$). Figure \ref{fig:originalTrajectory} depicts the trajectory $c$ associated with this scenario, showing projections of $c$ considering the variables \emph{carPosition}, \emph{battery}, \emph{brakes}, and \emph{lidarRange}. 

All 3 inputs for the process have now been defined: the design, the hazard and the scenario. Next, we generate the causal model.

\subsubsection{Building the causal model}

Our choice for elements that comprise the signature are shown below. Note that we depart from the choice made in Example~\ref{ex:end-ex-variables}, as to show that the set of endogenous variables can be chosen by the user. Recall that $\mathcal{U}$ is the set of exogenous variables and $\mathcal{V}$ is that of endogenous variables.

\begin{itemize}
    \item $\mathcal{U} =$ \{\emph{acceleration, pedestrianPosition, carPosition, g, critical, speed, decay,\\belowCritical}\}
    \item $\mathcal{V} =$ \{\emph{brakes, battery, lidarRange}\}
\end{itemize}

Our choice of endogenous variables is based not only on what can reasonably be viewed as a cause but, more importantly, is based on what the user has control over and is willing to analyse and change in order to fix the system. The goal for this particular analysis is to determine which parts of the car are related to the failure and should be modified so that the car avoids colliding with the pedestrian. 

Even though the pedestrian itself can be viewed as a part of the cause of the accident, we ultimately have no control over them, so we do not consider $pedestrianPosition$ as an endogenous variable. Furthermore, we also classify the car acceleration and speed as exogenous variables; in our example, the car is autonomous and always travels at the road speed until it detects an obstacle, in which case, the car slows down. These variables are part of the set of exogenous variables, i.e., the external setting that exerts influence on the system. 

Concerning the variables in the endogenous set, it includes the braking coefficient ($brakes$), the lidar range ($lidarRange$) and the battery level ($battery$). We aim to identify if they can be considered a cause for the collision with the pedestrian and, if applicable, the moment in time where the value of these variables can lead to the pedestrian being hit. As it happens, constants such as the braking coefficient can only be modified in its full trajectory. Variables, such as the battery, can change depending on which parts of the system is active, so it is possible to modify only a slice of it.


Initially, the set of trajectories that compose $\mathcal{R}$ is the trajectory ($c$) that led to the collision. As the search is conducted in the next steps, $\mathcal{R}$ will expand. The causal model is built using the sets of functions that relate the trajectories in $\mathcal{R}$ with the variables in $(\mathcal{U} \cup \mathcal{V})$. For example, considering the trajectory $\vec{u}$, we have:

\begin{itemize}
    \item $\mathcal{F}_{brakes}(...) = 0.2$
    \item $\mathcal{F}_{battery}(...) = 10 - (t*\textit{decay})$
    \item $\mathcal{F}_{lidarRange}(...) = \begin{cases} 
      20, & \textit{battery} \geq \textit{critical} \\
      10, & \textit{battery} < \textit{critical}
   \end{cases}$
\end{itemize}

Since there are no cyclic dependencies in our set of functions, the analysis can proceed.

\subsubsection{Searching for causes} 

We present, in this and in the next section (using the example), the algorithms both for the search (Algorithm~\ref{alg:causalitySearchAlgorithm}) and for determining causality (Algorithm~\ref{alg:causalityAlgorithm}).

\IncMargin{1em}
\begin{algorithm2e}

\SetKwInOut{Input}{input}
\SetKwInOut{Output}{output}

\SetKwProg{Fn}{Function}{ :}{end}
\SetKwFunction{Union}{Union}
\SetKwFunction{FindCompress}{FindCompress}
\SetKwFunction{Search}{Search}
\SetKwFunction{GetIntervals}{GetIntervals}
\SetKwFunction{Override}{Override}
\SetKwFunction{SearchHeuristic}{SearchHeuristic}
\SetKwFunction{IsCause}{IsCause}
\SetKwFunction{Add}{Add}
\SetKwFunction{IsEmpty}{IsEmpty}
\SetKwFunction{increaseGranularity}{increaseGranularity}
\SetKwFunction{Search}{Search}
\SetKwFunction{FocusSearch}{FocusSearch}
\SetKwFunction{IsWithin}{IsWithin}
\SetKwFunction{Evaluate}{Evaluate}
\SetKwFunction{SetMinus}{SetMinus}
\SetKwFunction{ChangedVariables}{ChangedVariables}
\SetKwFunction{GetSubsets}{GetSubsets}
\SetKwFunction{GetVariables}{GetVariables}
\SetKwFunction{GetSubsetsOfSizeOne}{GetSubsetsOfSizeOne}
\SetKwFunction{SatisfiesACOne}{SatisfiesACOne}
\SetKwFunction{SatisfiesACTwo}{SatisfiesACTwo}
\SetKwFunction{SatisfiesACThree}{SatisfiesACThree}
\SetKwFunction{Projection}{Projection}
\SetKwFunction{UpdateModel}{UpdateModel}
\SetKwFunction{Holds}{Holds}

\Input{\ CausalModel $M$;}
\Input{\ Trajectory $c, u$;}
\Input{\ BooleanPredicates $\Phi$;}
\Output{\ Set [(Trajectory, Trajectory, Set [Variable], TimeInterval)] $causes$;}

\Fn{\Search{$M$,$c$,$u$,$\Phi$}}{
    Integer $granularity$ = 2, $maximumGranularity$ = 10\;
    \ForEach {Set [Variable] $\vec{X}$ \textbf{in} \GetSubsetsOfSizeOne{$M.signature.V$}}{
        \ForEach {interval \textbf{in} \GetIntervals{$c$, $granularity$}}{
            (Set [Trajectory] $\vec{x}^\prime$, Set [Trajectory] $\vec{w}$) = \SearchHeuristic{$c$, $\vec{X}$, $interval$}\;
            Set [Variable] $\vec{W}$ = $\vec{w}$.\GetVariables{}\;
            \If {\IsCause{$M$, $u$, $\Phi$, $\vec{x}$, $\vec{x}^\prime$, $\vec{w}$, $\vec{X}$, $\vec{W}$}}{
                $causes$.\Add{($c$, $\vec{x}$, $interval$)}\;
            }
        }
    }
    \If {$granularity < maximumGranularity$}{
        \If {$causes$.\IsEmpty{}}{
            \increaseGranularity{$granularity$}\;
            \Search{}\;
        } \Else {
            \ForEach {$cause$ \textbf{in} $causes$}{
                \FocusSearch{$cause.interval$, $\vec{X}$}\;
            }
        }
    
    }
}

\Fn{\FocusSearch{$interval$, $\vec{X}$}}{
    \ForEach {$interval$ \textbf{in} \GetIntervals{$c$,$maximumGranularity$}}{
        \If{$interval$.\IsWithin{$interval$}}{
            (Set [Trajectory] $\vec{x}^\prime$, Set [Trajectory] $\vec{w}$) = \SearchHeuristic{$c$, $\vec{X}$, $interval$}\;
            Set [Variable] $\vec{W}$ = $\vec{w}$.\GetVariables{}\;
            \If {\IsCause{$M$, $u$, $\Phi$, $\vec{x}$, $\vec{x}^\prime$, $\vec{w}$, $\vec{X}$, $\vec{W}$}}{
                $causes$.\Add{($c$, $\vec{x}$)}\;
            }
        }    
    }
}


\caption{Pseudo-code for search algorithm.}
\label{alg:causalitySearchAlgorithm}
\end{algorithm2e}

The function iterates through each endogenous variable (line 3) and intervals of trajectory $c$ (line 4). In order to generate new trajectory slices, we apply a search heuristic (in this case, genetic algorithm \cite{mitchell1998introduction}) to find data points that will form the alternative trajectory slices for variables in the cause and in the contingency set (line 5). For that, we use the Global Optimisation toolbox \cite{global-optimisation-toolbox} for Matlab. The algorithm works on a population and attempts to find the global maxima or minima of a (fitness) function. A population is a set of points in the design space and it is initially generated randomly. The algorithm computes the next generation of the population by interacting with the fitness function and the individuals in the current generation until it finds the maxima. Our fitness function aims to generate slices that are diverse (in terms of distance and shape) from the slice that led to the fault.

Directly connecting the data points yielded by the search would result in an erratic curve. To mitigate this issue, in {\tt SearchHeuristic}, we employ a notion of curve fitting using a fourth degree polynomial equation as a smoothing function \cite{guest2012numerical}. This results in a smoother behaviour with the trade-off of the resulting curve being an approximation to the data points obtained by the search, instead of an exact match. As an alternative fit, one can use polynomial interpolation, which would result in a smooth curve that would exactly fit the points yielded by the search. However, this greatly increases computational costs since it requires functions with polynomial degree of (n-1), where n is the number of data points. The search typically yields thousands of data points, hence, for such large numbers, interpolation is impractical.

Once we have the alternative trajectories for variables in a prospective cause (sets $\vec{X}$ and $\vec{W}$), we call a function {\tt isCause} that will check for actual causality (the details of this function can be seen in Algorithm~\ref{alg:causalityAlgorithm}). If positive, then causes are added to the outputting set (line 8). We represent a cause as a tuple composed by the original hazardous trajectory $c$, the variables in the cause, and the time interval. 

After searching and determining causes, the algorithm attempts to increase granularity regardless of whether causes have been found or not. If a cause was found, we attempt to increase its precision (with respect to time) by searching the specific area where a cause has been found and, hence, trimming the time intervals into smaller pieces. Hence, if a cause is found with a granularity lower than the minimum, we focus the search on its specific time interval to increase precision (lines 27 to 40). 

Otherwise, if no cause is found for coarser time intervals, we gradually increase granularity (from 2 to 4 and finally 10, that is, each slice should represent 50$\%$, 25$\%$ and 10$\%$ of the trajectory ($c$) duration, respectively -- this is performed by the function $\mathit{increaseGranularity}$) and resume the search (lines 13 to 16). The reason for increasing granularity even though no cause is found is due to the following two reasons: 
\begin{itemize}

\item First, overriding only slices of a trajectory effectively turns it into a piecewise function. Since the degree of our polynomial function to smooth the curve is constant (regardless of how many data points there exist), the finer the time intervals the less data points fall within each time interval, and, thus, the easier it is for the smooth curve to exactly match the points resulting from the search. 

\item Second, because our search is non-exhaustive in nature,  the search can find and assess causality for more slices, if the time intervals are finer. This results in more causality assessments in overall and increases the likelihood to find causes.

\end{itemize}

There are two stopping criteria for the search. If the search does not find a cause even after splitting the trajectory into the maximum number of intervals (set to 10), then it stops. Otherwise, if it finds at least one cause, it focuses the search on those areas related to the causes in an attempt to increase precision (more subtle changes within shorter time intervals). After focusing (using the maximum number of time intervals), the search is stopped.

Once finished, another search is performed  considering the interaction between variables. Algorithm \ref{alg:causalitySearchAlgorithm} attempts to override one variable at a time. However, certain causes can only be found by overriding multiple variables at once. Thus, we also conduct the search considering pairwise and further considering sets of three variables. The pseudo-code for these additional searches are not shown here; in short, they lift the strategy presented in Algorithm~\ref{alg:causalitySearchAlgorithm} to consider multiple variables. Because the complexity of the search significantly increases, we do not attempt to search using sets with more than     3 variables.

\IncMargin{1em}
\begin{algorithm2e}
\SetKwInOut{Input}{input}
\SetKwInOut{Output}{output}
\SetKwProg{Fn}{Function}{ :}{end}

\Input{\ CausalModel $M$;}
\Input{\ Trajectory $c, u$;}
\Input{\ Set [Variable] $\vec{X}$;}
\Input{\ Set [Variable] $\vec{W}$;}
\Input{\ Set [Trajectory] $\vec{x}$;}
\Input{\ Set [Trajectory] $\vec{w}$;}
\Input{\ Set [Trajectory] $\vec{x}^\prime$;}
\Input{\ BooleanPredicates $\Phi$;}
\Output{\ Boolean $isCause$;}
\BlankLine

\Fn{\IsCause{$M$, $u$, $\Phi$, $\vec{x}$, $\vec{x}^\prime$, $\vec{w}$, $\vec{X}$, $\vec{W}$}}{
    $isCause$ = False\;
    \If{\SatisfiesACOne{$M$, $u$, $\Phi$}}{
        \If{\SatisfiesACTwo{$M$, $u$, $\Phi$, $\vec{x}^\prime$, $\vec{w}$, $\vec{X}$, $\vec{W}$}}{
            \If{\SatisfiesACThree{$M$, $u$, $\Phi$, $\vec{x}$}}{
                $isCause$ = True\;
            } 
        }     
    }
    \Return $isCause$\;
}

\Fn{\SatisfiesACOne{$M$, $u$, $\Phi$}}{
    \Return \Holds{$M$, $u$, $\Phi$}\;
}

\Fn{\SatisfiesACTwo{$M$, $u$, $\Phi$, $\vec{x}^\prime$, $\vec{w}$, $\vec{X}$, $\vec{W}$}}{
    $M_C^{updt}$ = \UpdateModel{$M$,\{($\vec{X}$, $\vec{x}^\prime$), ($\vec{W}$, $\vec{w}$)\}}\;
    $isAC2aTrue$ = \Holds{$M_C^{updt}$, $u$, $\neg \Phi$}\;
    \Return $isAC2aTrue$\;
}

\Fn{\SatisfiesACThree{$M$, $u$, $\Phi$, $\vec{x}$,}}{
   \ForEach {$subOfX$ \textbf{in} \GetSubsets{$\vec{x}$}}{

       ($\vec{x}^\prime$, $\vec{w}^\prime$) = \SearchHeuristic{$c$, $subOfX$, $interval$}\; $\vec{X}^\prime$ = $\vec{x}^\prime$.\GetVariables{}\;
       $\vec{W}^\prime$ = $\vec{w}^\prime$.\GetVariables{}\;

       \If{\SatisfiesACOne{$M$, $u$, $\Phi$}}{           
           \If{\SatisfiesACTwo{$M$, $u$, $\Phi$, $\vec{x}^\prime$, $\vec{w}^\prime$, $\vec{X}^\prime$, $\vec{W}^\prime$}}{
                \Return False\;
            } 
        }
    }
    \Return True\;
}

\caption{Pseudo-code for causal analysis algorithm.}\label{alg:causalityAlgorithm}
\end{algorithm2e}
\DecMargin{1em}

\subsubsection{Searching for a cause -- part \#1: lidar range}

To check if the set of slices is a cause, the method {\tt IsCause} asserts the causality clauses (as in Definition \ref{def:causeCPS}). The pseudo-code for this function can be seen in Algorithm~\ref{alg:causalityAlgorithm}. In this algorithm, we check the three clauses of Definition~\ref{def:causeCPS} (lines 3 to 5). Firstly, for AC1, the prospective cause ($\vec{x}$) is taken from $c$, which is built around causal model $M$ in Algorithm~\ref{alg:causalitySearchAlgorithm} (and, thus, $(M,u) \models (c \mapsfrom \vec{x})$). Therefor, for AC1, we only need to check if the effect holds in the causal model ($(M,u) \models \Phi$). As for AC2, we check whether the slices found by the search suffice to violate $\Phi$ (lines 16 to 18). Finally, for AC3, we check whether any subset of the trajectory slices (line 22) satisfy AC1 and AC2 (lines 25 and 26). To do this, we perform a quicker version of the search (focusing on the subsets of $\vec{x}$), looking for alternative slices for $\vec{X}$ and $\vec{W}$ that satisfy both AC1 and AC2. We have previously noted that, in Definition~\ref{def:causeCPS}, the set $\vec{w}$ is a subset of $\mathrm{Trajs}(\vec{W}) \cap \mathcal{R}$ and hence, we do not need to check for all possible variations in $\mathrm{Trajs}(\vec{W})$, but only the ones that are also in the set $\mathcal{R}$, which is built iteratively by the search. We discuss this algorithm in more details (including its soundness) in Appendix~\ref{sec:soundness}.

Considering Definition \ref{def:causeCPS}, we have the previously defined signature ($S_C$), the causal model $M$, the trajectory $c$ and $u = c\downarrow_{\mathcal{U}}$. The search is initiated and, in its first phase, it considers variables individually. A potential cause is identified when, after a trajectory slice is overridden, $\Phi$ does not hold. 

\begin{figure}[!h]
  \centering
  \begin{subfigure}[b]{0.3\linewidth}
    \centering
    \includegraphics[width=1\linewidth]{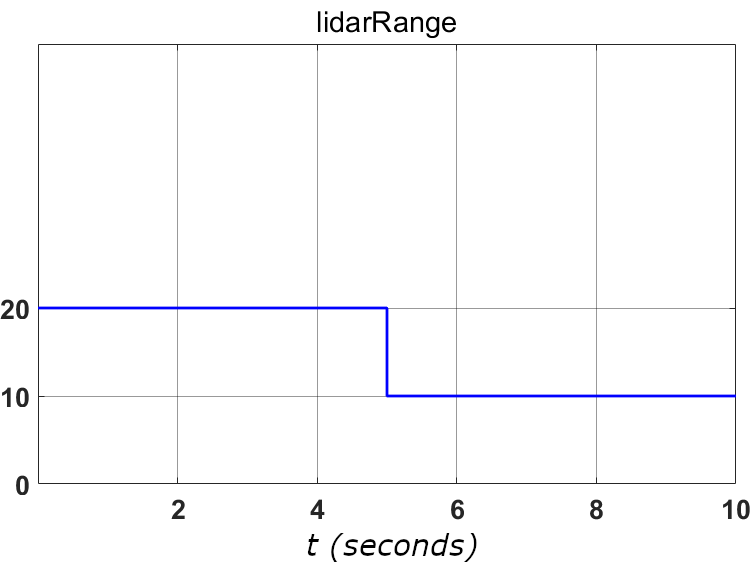} 
    \caption{$x$.} 
    \label{fig:x_X} 
  \end{subfigure}
  \begin{subfigure}[b]{0.30\linewidth}
    \centering
    \includegraphics[width=1\linewidth]{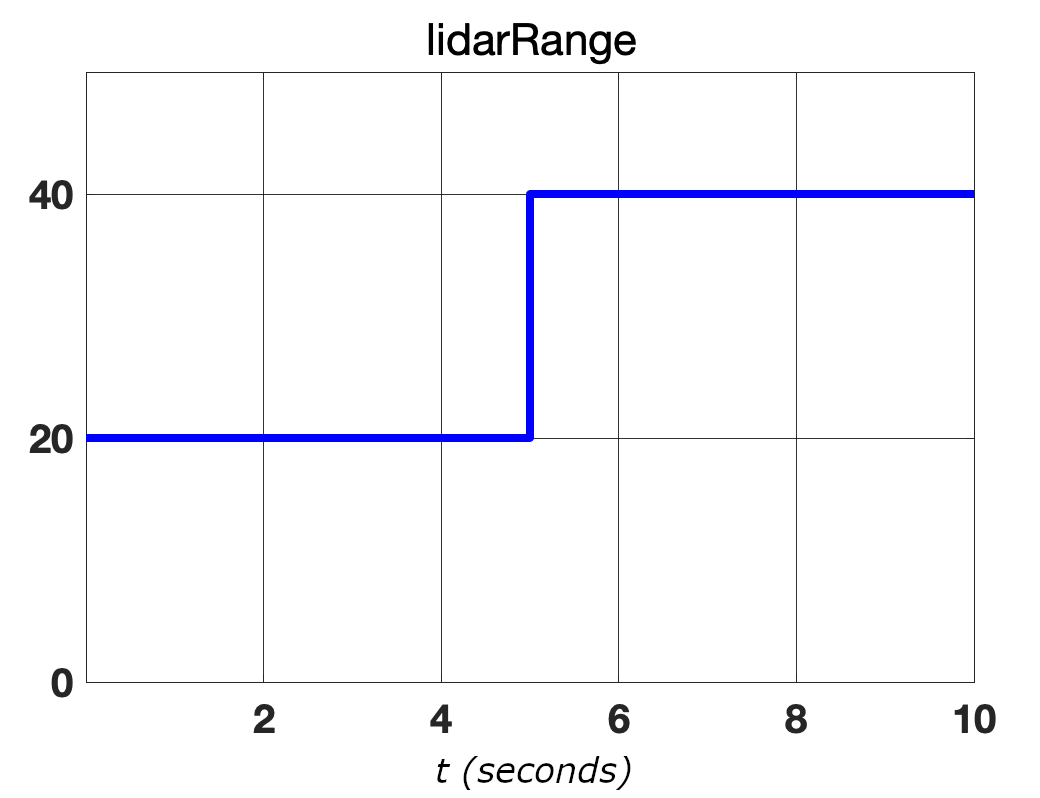} 
    \caption{$x^\prime$.} 
    \label{fig:y_X} 
  \end{subfigure}
  \begin{subfigure}[b]{0.30\linewidth}
    \centering
    \includegraphics[width=1\linewidth]{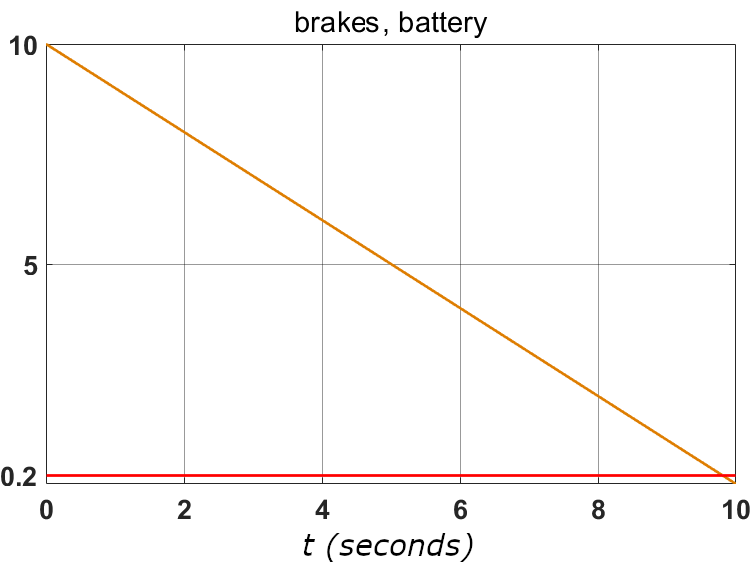} 
    \caption{$w$.} 
    \label{fig:w_W} 
  \end{subfigure}
  \caption{Trajectory slices related to the cause.}
  \label{fig7}
\end{figure}

One of the causes presented is related to the range of the lidar. In the context of the theory, we have that $\Vec{X} = \{lidarRange\}$. Figure \ref{fig:x_X} depicts the original \emph{lidarRange} trajectory, i.e., $\vec{x} = \{x \downarrow_{lidarRange}\}$. Since the duration of the system simulation (i.e., the duration of trajectory $c$) is 10 seconds, the heuristic considers time slices of 5\emph{s}, 2.5\emph{s}, and 1\emph{s}, as explained in Section \ref{ssec:process}. Going back to Definition \ref{def:causeCPS}, our algorithm assesses whether $(c \mapsfrom \vec{x})$ is a cause of $\Phi$. For that, the 3 clauses associated with this definition must be satisfied.

The trajectory $c$ and the set of slices $\vec{x}$ lead to $\Phi$, as shown in Figure \ref{fig:originalTrajectory}, and, thus, $AC1$ is satisfied. Additionally, $\vec{x}$ is a singleton, which trivially satisfies $AC3$.

\begin{figure}[!h]
  \centering
  \includegraphics[width=0.5\linewidth]{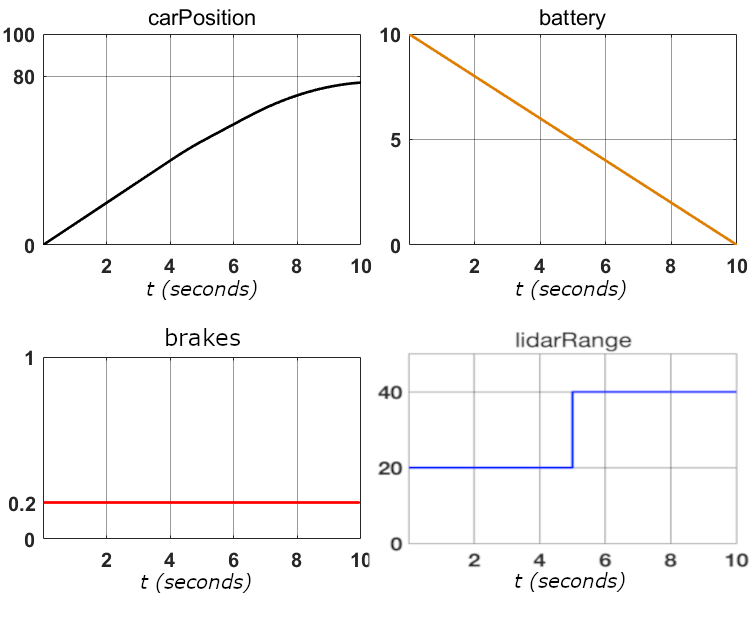} 
  \caption{$\vec{X} \leftarrow \vec{x}^\prime, \vec{W} \leftarrow \vec{w}$.} 
  \label{fig:M_yX} 
\end{figure}

For $AC2$, we consider the set $\vec{W} = \{battery, brakes\}$, and the sets of trajectories $\vec{x^\prime} = \{x^\prime \downarrow_{lidarRange}\}$ (Figure \ref{fig:y_X}) and $\vec{w} = \{w \downarrow_{battery},w \downarrow_{brakes}\}$ (Figure \ref{fig:w_W}). We can see that $\Phi$ does not hold for $(M_{\vec{X} \leftarrow \vec{x}^\prime, \vec{W} \leftarrow w},  u)$ (see Figure \ref{fig:M_yX}). Thus, $AC2$ is satisfied.

Increasing the range of the lidar alone would suffice to prevent the collision. The reason is that the car can detect the pedestrian at a greater distance. Although it is a valid cause, the search continues, now considering the remainder of the endogenous variables. Our search presents to the user every cause found. For instance, one could argue that it could be prohibitively expensive to install a much more powerful lidar in an actual car.


\subsubsection{Searching for a cause -- part \#2: battery and brakes separately}

No additional causes have been found whilst the search considered these variables individually. As an illustration, we present some of the trajectories that were considered by the search. 

\begin{figure}[!ht]
  \centering
  \begin{subfigure}[b]{0.49\linewidth}
    \centering
    \includegraphics[width=1\linewidth]{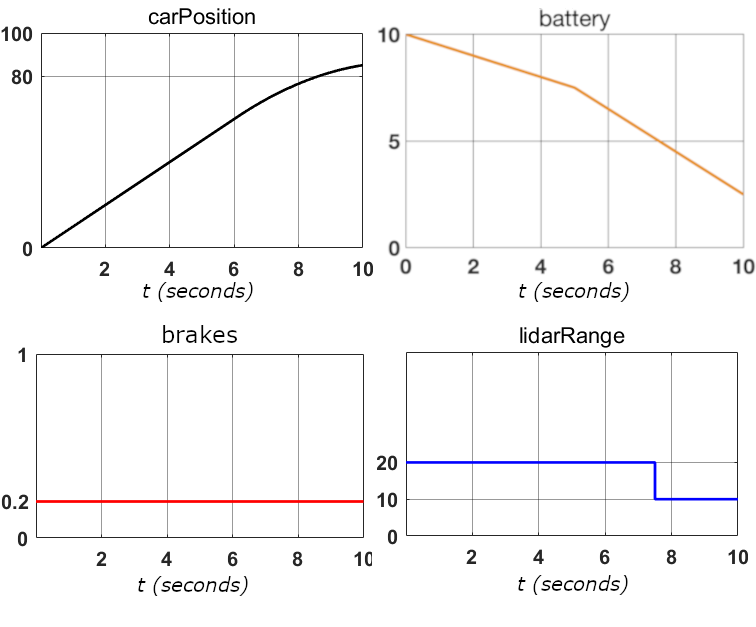} 
    \caption{Overriding only the battery.} 
    \label{fig:overridingBattery} 
  \end{subfigure}
  \begin{subfigure}[b]{0.49\linewidth}
    \centering
    \includegraphics[width=1\linewidth]{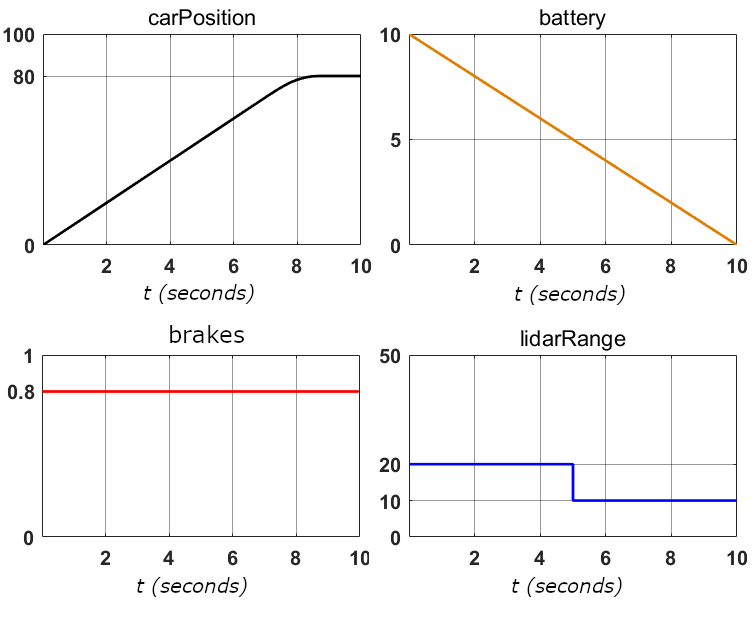} 
    \caption{Overriding only the brakes.} 
    \label{fig:overridingBrakes} 
  \end{subfigure}
  \caption{Overriding the trajectories (search -- part \#2).}
  \label{fig7c}
\end{figure}

For instance, the battery was overridden during the time interval $[0,5)$, depicted in Figure \ref{fig:overridingBattery}. Due to an improved battery, the lidar works in long range mode, detects the pedestrian and starts braking sooner. However, with the default braking system the car still hits the pedestrian. Thus, the battery alone has not been identified as a cause. 

Analogously, the search considered the braking system and overrode it to the highest value permissible by the design. This resulted in a much lower braking distance, as shown in Figure \ref{fig:overridingBrakes}. However, this also does not prevent the accident. The car brakes harder but too late since the lidar is in a lower range mode due to a battery that is below the critical threshold. The brakes alone have also not been identified as a cause.

So far, when analysing the endogenous variables individually, only the lidar range has been identified as a cause. The next step is to consider variables pairwise, that is, overriding two variables at the same time.

\subsubsection{Searching for a cause -- part \#3: battery and brakes simultaneously}

Since the lidar range has been identified as a cause by itself, the search ignores this variable when searching for causes related to multiple variables. The reason for that is the minimality clause, AC3.

\begin{figure}[h]
\centering
\includegraphics[width=0.5\linewidth]{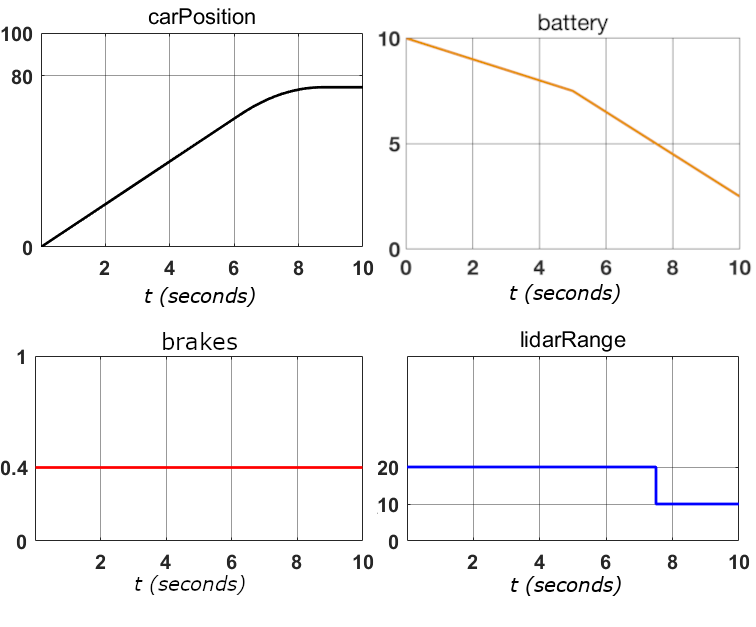} 
\caption{Overriding both battery and braking coefficient (search -- part \#3).} 
\label{fig:overridingBoth} 
\end{figure}

A second cause was found when overriding trajectory slices for both the battery and the brakes. As shown in Figure \ref{fig:overridingBoth}, the car brakes harder and sooner, thus avoiding the accident. As a conclusion, the slices for both the battery and the brakes together can also be identified as a cause for the collision.

We emphasise that some limitations apply to the variables in the interventions. For example, its maximum and minimum values and whether the variable is a constant. This is to avoid finding causes that do not respect physics or system requirements. In this case study, reasonable values for the braking coefficient is between 0.2 and 0.8 and, thus, having a braking coefficient of, for instance, 300 would generate a cause that does not respect real world scenarios. These restrictions can be seen as applying restrictions to the possible trajectories in $\mathcal{R}$.
\section{Empirical evaluation}
\label{sec:caseStudies}

In this section, we present two case studies and a series of benchmarks that explore the applicability of our technique for systems more complex than the running example.\footnote{A lab package for the experiments described in this section can be found on Zenodo:\\ \url{http://tinyurl.com/CausalityForCPS-labpackage}} More precisely, we describe the application of the strategy proposed in Section \ref{ssec:process} to two case studies. In the first case study, we find causes for (systematically inserted) faults on a suspension system of a vehicle. In the second one, we explore an autonomous vehicle platoon to find causes behind communication issues. For the benchmarks, we explore 4 scalable systems.

\subsection{Research objectives}
\label{research-questions}

The main goal of the case studies is the evaluation of our causal analysis technique in the context of two verification problems. The first one is about whether it can be applied to identify the causes of (inserted) faults in the selected CPS. As for the second problem, we aim to assess whether it can also identify causes in hazardous situations due to design oversights. In particular, the case studies illustrate how the causality analysis can be used to identify injected faults and design oversights.

We consider case studies containing failures and hazards reported in previous works \cite{araujo2019multiobjective, araujo2020connected}, and apply our technique by defining the effects ($\Phi$), building the causal models, and assessing causality.

\subsection{Case study \texorpdfstring{$\#$1:} \\ verifying a suspension system}
\label{caseStudyPlatooning}

\begin{figure}[!h]
	\centering
    \begin{subfigure}[b]{0.5\textwidth}
    	\includegraphics[width=230pt]{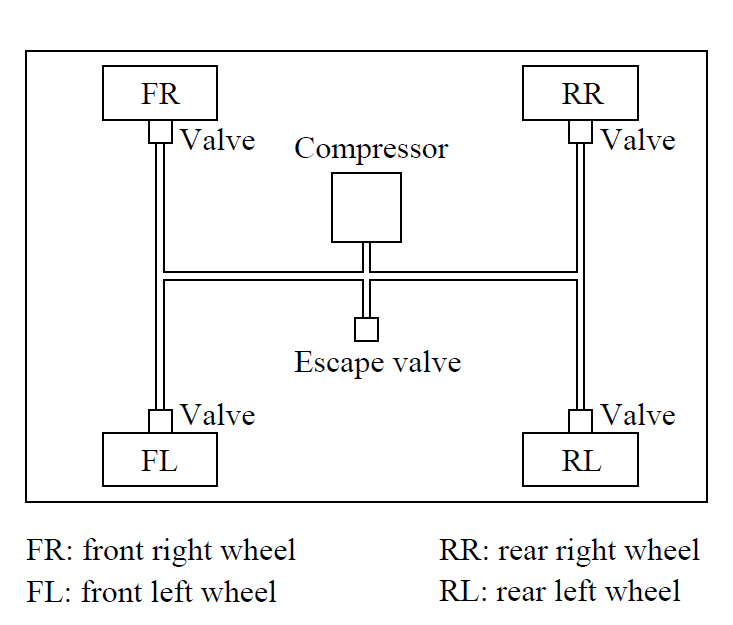}
        \caption{Suspension system overview.}
    	\label{fig:suspension}
    \end{subfigure}
    \hspace*{\fill}
    \begin{subfigure}[b]{0.3\textwidth}
    	\includegraphics[height=190pt]{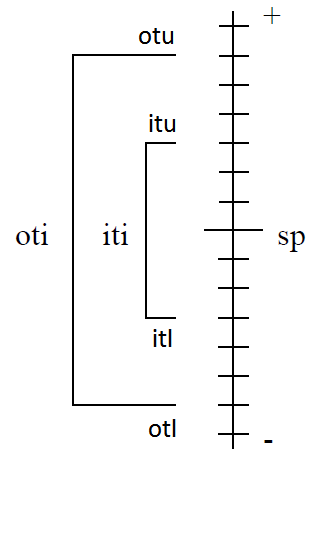}
        \caption{Tolerance levels.}
    	\label{fig:suspensionTolerance}
	\end{subfigure}
  	\caption{Suspension system.}
    \label{fig:suspensionSystem}
\end{figure}

In this first case study, we examine an automotive pneumatic suspension system \cite{muller2000modelling}. We make use of results from a previous experiment \cite{araujo2019multiobjective}, in which the faults were manually inserted in the system via a mutation process and detected using Hyconf. The purpose here is to confirm whether our analysis can detect their causes and assist with the correction of such faults. The system's goal is to increase driving comfort by adjusting the chassis level to compensate for road disturbances. This is achieved by a  suspension system that connects the valves attached to each wheel to a compressor and an escape valve (see Figure \ref{fig:suspension}).

The system aims to keep the chassis level as close as possible to a defined set point in each of the four wheels. The decision to increase or decrease the chassis level is based on the tolerance intervals defined for each wheel, as depicted in Figure \ref{fig:suspensionTolerance}. We consider $[sp - otl, sp + otu]$ and $[sp - itl, sp + itu]$ as the outer and inner tolerance intervals, respectively. Here, $sp$ represents the set point, which is the target value of the chassis level, and \emph{itu, itl, otu} and \emph{otl} represent the inner and outer tolerance thresholds along with their respective upper and lower values. Table~\ref{tab:ss_inputs} displays the variables in the system.

\begin{table}[!h]
\centering
\caption{Suspension system variables.}
\label{tab:ss_inputs}
\begin{tabular}{|l|c|c|c|}
\hline
\rowcolor[HTML]{EFEFEF} 
\multicolumn{1}{|l|}{\cellcolor[HTML]{EFEFEF}\textbf{Name}} & \multicolumn{1}{|c|}{\cellcolor[HTML]{EFEFEF}\textbf{Type}} & \multicolumn{1}{|c|}{\cellcolor[HTML]{EFEFEF}\textbf{Description}} \\ \hline

\textit{dist} & Input & Road disturbance  \\ \hline
\textit{cp} & Input & Compression valve \\ \hline

\textit{ev} & Input & Escape valve \\ \hline
\textit{bend} & Input & Car is on a bend \\ \hline
\textit{c} & Output & Influence of valves on the chassis \\ \hline
\textit{h} & Internal & Current chassis level \\ \hline
\textit{f} & Internal & Filtered chassis level \\ \hline
\textit{e} & Internal & Filtered disturbance \\ \hline
\textit{T} & Internal & Filter constant \\ \hline
\textit{sp} & Internal & Set point of the chassis \\ \hline
\textit{otu} & Internal & Outer tolerance upper limit \\ \hline
\textit{otl} & Internal & Outer tolerance lower limit \\ \hline
\textit{itu} & Internal & Inner tolerance upper limit \\ \hline
\textit{itl} & Internal & Inner tolerance lower limit\\ \hline

\end{tabular}
\end{table}

The system receives four inputs and it outputs the current chassis level $h$. The inputs are $dist$, $cp$, $ev$, and $bend$. The first one (\emph{dist}) corresponds to the disturbance level coming from the environment, which indicates road perturbations such as small depressions or elevations. The \emph{cp} and \emph{ev} inputs dictate the change in the chassis level performed by the compressor and the escape valve, respectively. Finally, the variable $bend$ indicates whether the vehicle is turning, which prohibits the adjustment of the pneumatic levels.

The control flow is described as follows. The system starts with the chassis set within the tolerance interval and all valves, as well as the compressor, are closed. Changes to the chassis level are constantly being monitored in order to determine the need to its increase or decrease. This is done by comparing the filtered chassis level ($f$) to tolerance limits ($otu$, $otl$, $itu$, $itl$). The filtered chassis value is obtained by setting is first derivative to an equation that depends on the current chassis level, the previous filtered value and the filter constant ($\dot{f} = \frac{h - f}{T}$). On the other hand, the current chassis level $h$ depends on the influence of the compressor and escape vales and on the filtered road disturbances ($\dot{h} = \dot{c} + \dot{e}$).

In what follows, we manually insert faults to the actual implementation and apply our process to determine their causes. 

\subsubsection{Defining the faults}

We have selected two mutants from our previous experiment \cite{araujo2019multiobjective} and an overview of them can be seen in Table \ref{tab:suspensionfaults}. A manual inspection guarantees that the mutants are non-equivalent. The first mutant changes the value of a system constant. The second replaces the value of a variable when the system should be lowering the chassis.

\begin{table}[!h]
    \centering
    \caption{Faults inserted and the issue observed.}
	\label{tab:suspensionfaults}
    \resizebox{1\textwidth}{!}{%
	\begin{tabular}{|c|c|c|c|}
		\hline
		\cellcolor{Gray} \textbf{Original} & \cellcolor{Gray} \textbf{Mutation} & \cellcolor{Gray} \textbf{Effect} & \cellcolor{Gray} \textbf{$\Phi$} \\ \hline
			$otu = 5$ & $otu = 6$ & Pressure increases more than allowed around 10s. &$c \downarrow_{\{f\}} >_{[8,12)} c \downarrow_{\{sp\}} + 5$ \\ \hline
			$\dot{e} = dist$ & $\dot{e} = 0$ & Sometimes, disturbances do not affect the system. & $c \downarrow_{\{dist\}} >_{\mathrm{dom}(c)} 0 \wedge c \downarrow_{\{c\}} =_{\mathrm{dom}(c)} 0$ \\ \hline
	\end{tabular}%
    }
\end{table}

\subsubsection{Building the causal model}

\bigskip

Here, we define the signature by selecting the endogenous and exogenous variables, in addition to building the causal model to assess the failures. With respect to the first fault, the noticed effect is a high pressure in the tires. As for the second fault, the issue is that road disturbances are not affecting the system sometimes. To properly fix these faults, we need to identify why and when they happen.




Considering we are trying to find causes for system failures, we do not want to assign inputs as causes for such faults. Inputs are external factors and cannot be controlled by the system. Thus, we select as endogenous variables everything but the inputs $dist$, $cp$, $ev$ and $bend$.

\begin{itemize}
    \item $\mathcal{V} =$ \{\textit{e, c, h, f, T, otu, itu, itl, otl, sp}\}
    \item $\mathcal{U} =$ \{\textit{dist, cp, ev, bend}\}
\end{itemize}







With respect to $\mathcal{R}$, we start with the trajectory that led to the issue in each case and more trajectories are added to each $\mathcal{R}$ as the search is conducted. The initial trajectories were found in a previous experiment during conformance verification \cite{araujo2019multiobjective}. The structural equations are as follows.

\begin{itemize}
    \item $\mathcal{F}_{e}(...) = t * dist$
    \item $\mathcal{F}_{c}(...) =
    \begin{cases} 
      6.8 + t * ev, & f > sp + otu \\
      6.8 + t * cp, & f < sp + otl \\
      6.8 + t * 0, & otherwise
   \end{cases}$
    \item $\mathcal{F}_{h}(...) = t * (\dot{e} + \dot{c})$
    \item $\mathcal{F}_{f}(...) = t * ((h-f)/T)$
    \item $\mathcal{F}_{T}(...) = 10.3$
    \item $\mathcal{F}_{otu}(...) = 5$
    \item $\mathcal{F}_{itu}(...) = 4$
    \item $\mathcal{F}_{itl}(...) = 2$
    \item $\mathcal{F}_{otl}(...) = 1$
    \item $\mathcal{F}_{sp}(...) = dist + ((itl + itu) / 2) * 1/T$
\end{itemize}

As defined in Section \ref{sec:theoryContinuous}, for certain slices to be classified as cause, 3 clauses need to be satisfied. In our implementation, AC1 and AC3 are automatically satisfied as they always hold by construction. First, the search only tries to override trajectory slices that led to the fault, considering the causal model, thus AC1. Second, the search starts looking at slices individually before expanding the search to consider multiple variables. Furthermore, we do not consider the variables that have already been included in causes when considering further combinations, thus AC3. In what follows, we will focus our analysis on clause AC2. 

\subsubsection{Determining cause -- mutant \#1}

\begin{figure}[!h]
  \centering
  \begin{subfigure}[b]{0.49\linewidth}
    \centering
    \includegraphics[width=1\linewidth]{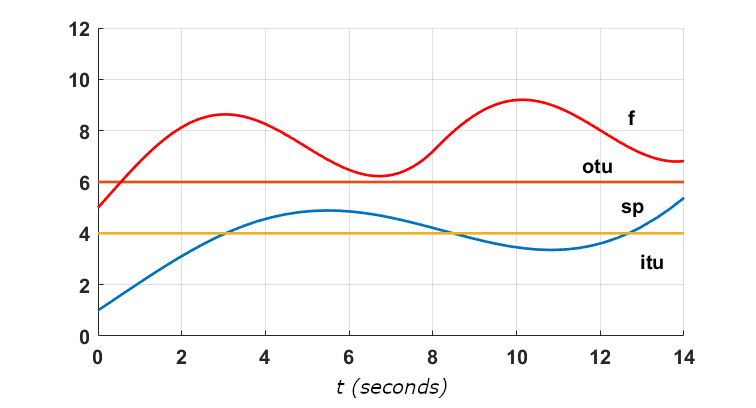} 
    \caption{Trajectory $c$ that led to the first failure.} 
    \label{fig:suspension1} 
  \end{subfigure}
  \begin{subfigure}[b]{0.49\linewidth}
    \centering
    \includegraphics[width=1\linewidth]{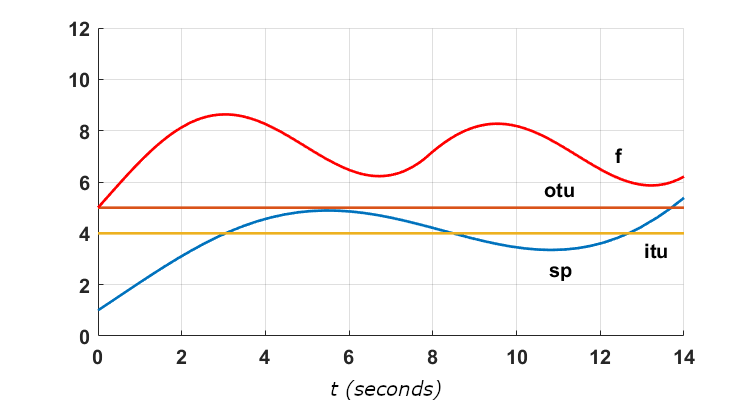} 
    \caption{Overriding $otu$.} 
    \label{fig:suspension1_fix} 
  \end{subfigure}
  \caption{Causal analysis for the first mutant.}
  \label{fig:suspension-mutant-1}
\end{figure}

In this step, the search looks for trajectory slices candidates and confirms whether they are actual causes. The first failure is observed when the chassis level increases further than what is expected; the system should lower the pressure when $f > sp + 5$, which does not happen immediately in the implementation (see Figure~\ref{fig:suspension1}): When $t \approx$ 10 $s$, we have that $sp \approx 3.71$ and $f \approx 9.27$.

The cause (i.e., in $c$, the variable $otu$ is equals to 6 during the time interval [0,14)) is trivially found by the search, when it assesses whether $(c \mapsfrom  \vec{x})$ when $\vec{x} = \{x_{otu}\}$. The failure ceases to occur when the constant $otu$ is reduced (see Figure~\ref{fig:suspension1_fix}), which is a clear indication that $otu$ might have been set to a higher value than it was required.

\subsubsection{Determining cause -- mutant \#2}

\begin{figure}[!h]
  \centering
  \begin{subfigure}[b]{0.49\linewidth}
    \centering
    \includegraphics[width=1\linewidth]{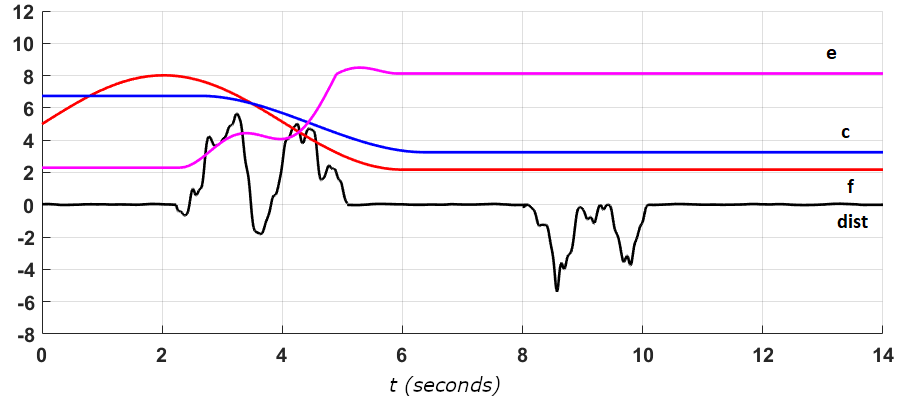} 
    \caption{Trajectory $c$ that led to the second failure.} 
    \label{fig:suspension2} 
  \end{subfigure}
  \begin{subfigure}[b]{0.49\linewidth}
    \centering
    \includegraphics[width=1\linewidth]{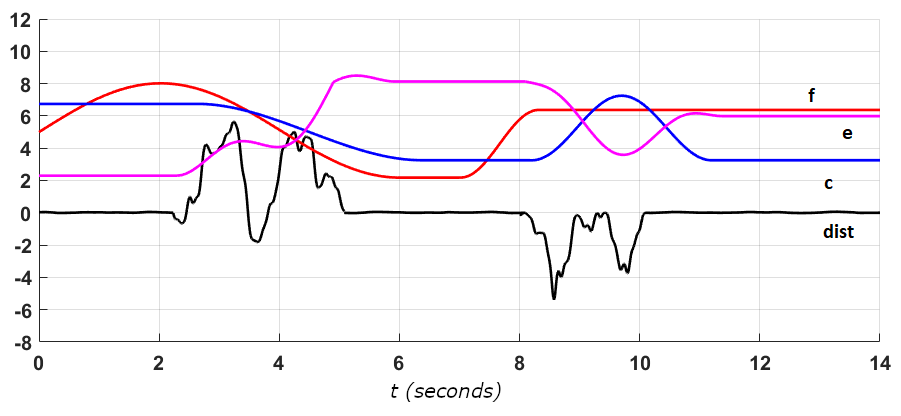} 
    \caption{Overriding $e$ and $f$.} 
    \label{fig:suspension2_fix} 
  \end{subfigure}
  \caption{Causal analysis for the second mutant.}
  \label{fig:suspension-mutant-2}
\end{figure}

With respect to the second fault, to find its cause it was necessary to override 2 slices of the original problematic trajectory. The trajectory displaying the issue is depicted in Figure \ref{fig:suspension2}. The first disturbance affects the system and it responds accordingly: the chassis level is lowered and so is the value of $f$. The second disturbance, however, does not affect the system.


The search found that no slices of $e$ or $f$ alone have been identified as cause. To find a situation where $\neg \Phi$, it was necessary to override two slices, as shown in Figure \ref{fig:suspension2_fix}. Firstly, the filtered chassis level  $f$ was overridden and increased in the $[7,8)$ time interval, and, secondly, the value of $e$ had to be changed during the $[8,10)$ time interval, which triggered a correction on the chassis level $c$. The AC2 clause was only satisfied when these two slices were overridden together.

This cause seems intuitive. Due to the fact that if the system is not affected by $dist$ but is affect by $e$, we can assume that there is a problem in the disturbance equation. Moreover, the search increased $f$ to the point that the system left the $down$ location, which suggests that the problematic equation is associated with this particular location.

\subsection{Case study \texorpdfstring{$\#$2:} \ \  exploring the design of a connected platoon}
\label{sec:caseStudyPlatooning}

Vehicular platooning is a cooperative and autonomous driving technology for linking two or more vehicles in a convoy. The goal of the convoy is to keep a close but safe distance between the vehicles using V2V (vehicle-to-vehicle) communications and automated driving technologies~\cite{bergenhem2012overview}. 

We have built a model of such a system \cite{araujo2020connected} whose goal is that all vehicles in the platoon should keep a safe distance but within each other’s communications range. In this model, the leading vehicle is driven by a human driver, while the velocity of the following vehicles is autonomously controlled; the autonomous followers should keep up with the leading vehicle's velocity.


The communication rules follow the standard defined in the ETSI EN 302 637-2 (Cooperative Awareness Basic Service - CAS) documentation \cite{ETSI302637}, which, among others, describes the rules for the frequency of packet transmission. These packets comprise Cooperative Awareness Messages (CAM), which contain information about the vehicle, such as acceleration, direction and position. Rules for sending a packet are parameterised and take into consideration three factors: (i) whether a vehicles has moved a long enough distance, (ii) a certain amount of time has passed, and (iii) its speed has changed above a certain threshold. If any of the three cases holds, then the vehicle must send a packet to communicate with the others. Moreover, we use a simple controller called the Intelligent Driver Model (IDM) \cite{treiber2000congested} in order to accelerate and decelerate the followers.

\begin{figure}[!h]
	\centering
	\includegraphics[width=1\textwidth]{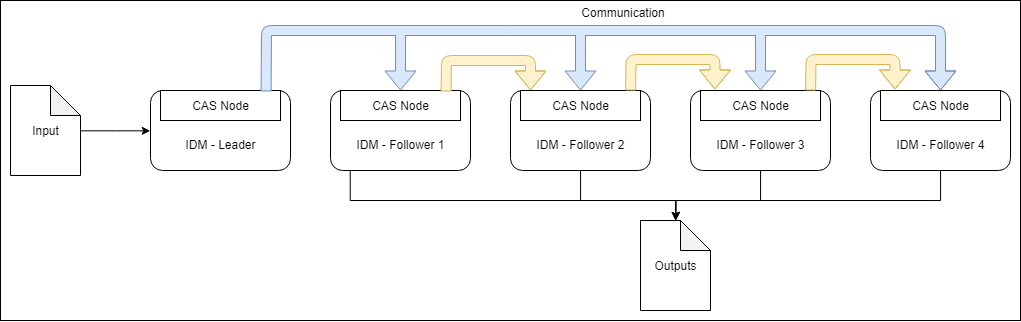}
	\caption{Platooning communication diagram \cite{araujo2020connected}.}
	\label{fig:platooningDiagram}
\end{figure}

An overview of our model can be seen in Figure  \ref{fig:platooningDiagram}. In the communication architecture that we use, each vehicle receives information from the leader and from the vehicle in front of it. This allows for each follower to learn about the leader's manoeuvres soon after they happen and anticipate them before they are fully propagated through the platoon. Other design decisions regarding the communication architecture may lead to different analysis results; for instance, using direct communication with the remaining followers might increase safety but also channel congestion.

The current model only takes longitudinal movement of the vehicle into account. We assume that the platoon moves along a very long highway without any drastic changes in direction. The main input of our testing campaign is the behaviour of the human driver in the lead vehicle (i.e., its acceleration). Once this input is generated and provided to the lead vehicle, then, via V2V communication, the convoy of followers must autonomously adjust their behaviour to match the vehicle in front.  The other inputs to our algorithms are the design-space parameters of the CAS protocol: by automatically searching through the design space, we explore the effect of these parameters on the safety and the quality of the platoon behaviour.

\subsubsection{Causal analysis}

In a previous work \cite{araujo2020connected}, we used HyConf to analyse how our model fared with respect to the ETSI standard. For that, we conducted a controlled experiment to compare the default parameters against alternative values. In order to generate inputs for continuous systems, a search-based approach was used: we formulated a multi-objective search problem that maximises hazard likelihood, data age as well as coverage of the input space via diversity of test inputs. Our approach automatically generated scenarios that resulted in hazardous situations (i.e., collision) whilst abiding by the standard. 

\begin{figure}[!h]
    \centering
    \includegraphics[width=0.8\linewidth]{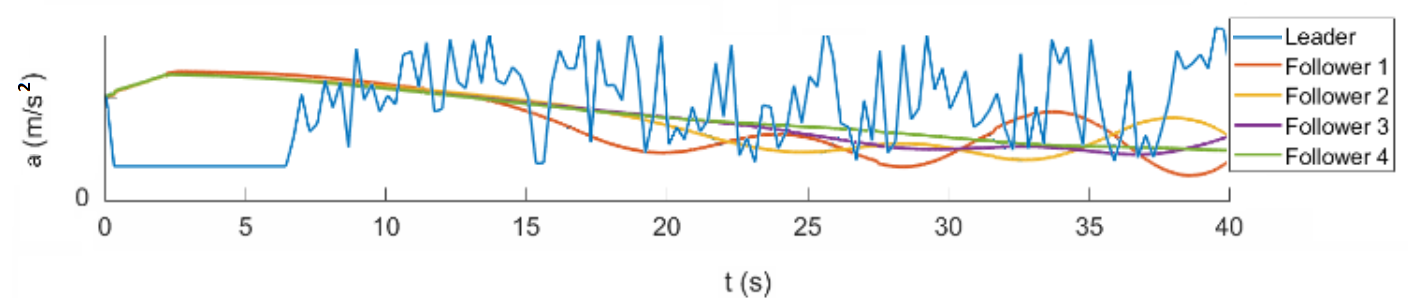}
    \caption{Example of a scenario generated by HyConf \cite{araujo2020connected}.}
    \label{fig:platooning_input}
\end{figure}

Here, we select the faulty scenarios obtained by Araujo et al. \cite{araujo2020connected} and use our causal analysis to explore safety levels between network parameters and platoon speeds. The scenarios are a combination of variations in the parameters of the ETSI standard to trigger CAMs (see Table~\ref{tab:ETSIparameters}) as well as patterns of acceleration and deceleration from the leading vehicle generated by HyConf (see Figure~\ref{fig:platooning_input}). 

\begin{table}[!h]
    \centering
    \caption{ETSI 302 637-2 CAM triggering parameters (adapted from \cite{araujo2020connected}).}
	\label{tab:ETSIparameters}
	\begin{tabular}{c|c|c|c|c|}
		\hhline{|~|-|-|-|-|}
		\multicolumn{1}{c|}{} & \cellcolor{Light_Gray} $T_{\min}$ & \cellcolor{Light_Gray} $T_{\max}$ & \cellcolor{Light_Gray} $d_{\min}$ & \cellcolor{Light_Gray} $v_{\min}$ \\ \hline
		\multicolumn{1}{|c|}{Default} & 100 $Hz$ & 1000 $Hz$ & 4 $m$ & 0.5 $m/s$ \\\hline
	    \multicolumn{1}{|c|}{Increased frequency} & 50 $Hz$ & 500 $Hz$ & 2 $m$ & 0.25 $m/s$ \\\hline
		\multicolumn{1}{|c|}{Decreased frequency} & 125 $Hz$ & 1250 $Hz$ & 5 $m$ & 0.625 $m/s$ \\\hline
	\end{tabular}
\end{table}

Table~\ref{tab:ETSIparameters} shows the default values for the parameters as well as two variations that increase or decrease the overall frequency rate. The parameters are for the minimum and maximum frequency ($T_{min}$ and $T_{max}$, respectively), minimum covered distance ($d_{min}$) and minimum change in speed ($v_{min}$). Furthermore,  Figure~\ref{fig:platooning_input} depicts one of the acceleration scenarios generated as input that led to a faulty behaviour. 

We conduct a causal analysis on two incidents. For each one of them, we define the corresponding $\Phi$ that describes the failure scenarios. 

\begin{table}[!h]
    \centering
    \caption{Observed failures.}
	\label{tab:platooningfaults}
    \resizebox{1\textwidth}{!}{%
	\begin{tabular}{|c|c|c|}
		\hline
		\cellcolor{Gray} \textbf{Variation} & \cellcolor{Gray} \textbf{Effect} & \cellcolor{Gray} \textbf{$\Phi$} \\ \hline
			Increased frequency & Follower 4 goes out of range (60 $m$) & $c \downarrow_{\{f4.position\}} + 60 <_{\mathrm{dom}(c)} c \downarrow_{\{f3.position\}}$  \\ \hline
			Decreased frequency & Follower 2 collides with follower 1 & $c \downarrow_{\{f2.position\}} \geq_{\mathrm{dom}(c)} c \downarrow_{\{f1.position\}}$  \\ \hline
	\end{tabular}%
    }
\end{table}

In this case, we would like to assess whether the vehicles kinematic rules can be interpreted as cause of the faulty behaviour instead of the network parameters. For simplicity, we classify the endogenous variables of the system as the leader and the followers, and the remaining variables are classified as exogenous. The leading vehicle has its own acceleration, speed and position (the variables are named $leader.acceleration$, $leader.speed$ and $leader.position$, respectively). Furthermore, the exogenous variable $u_1$ captures the leader acceleration (which is a abstraction of the car pedal position, wind speed, friction, tire conditions) and serves as input to the system. This variable is then linked to $leader.acceleration$. The remaining exogenous variables are related to communication rates.

\begin{itemize}
    \item $\mathcal{V} =$ \{\textit{leader, f1, f2, f3, f4}\}
    \item $\mathcal{U} =$ \{\textit{$T_{min}$, $T_{max}$, $v_{min}$, $d_{min}, u1$}\}
\end{itemize}

Furthermore, the structural equations are defined as follows. For simplicity, we choose to omit the equations for the second, third and fourth follower, but they follow the same pattern as the equations for the first follower.

\begin{itemize}
    \item $\mathcal{F}_{leader.acceleration}(...) = u_1$
    \item $\mathcal{F}_{leader.speed}(...) = 22.2 + leader.acceleration * t $
    \item $\mathcal{F}_{leader.position}(...) = 0 + (22.2*t + 0.5 * leader.acceleration * t^2)$
    
    \item $\mathcal{F}_{f1.acceleration}(...) = a_{IDM}(leader.position, f1.speed, \mid f1.speed - v_0\mid)$
    \item $\mathcal{F}_{f1.speed}(...) = 22.2 + f1.acceleration * t $
    \item $\mathcal{F}_{f1.position}(...) = 0 + (22.2*t + 0.5 * f1.acceleration * t^2)$
    \item $\mathcal{F}_{f1.v_0}(...) = 22.2 $
    \item $\mathcal{F}_{f1.\delta}(...) = 4$
    \item $\mathcal{F}_{f1.T}(...) = 1.5$
    \item $\mathcal{F}_{f1.s_0}(...) = 2$
    \item $\mathcal{F}_{f1.a}(...) = 4$
    \item $\mathcal{F}_{f1.b}(...) = 3$
\end{itemize}

\subsubsection{Results}

\begin{figure}[!h]
    \centering
    \includegraphics[width=0.8\linewidth]{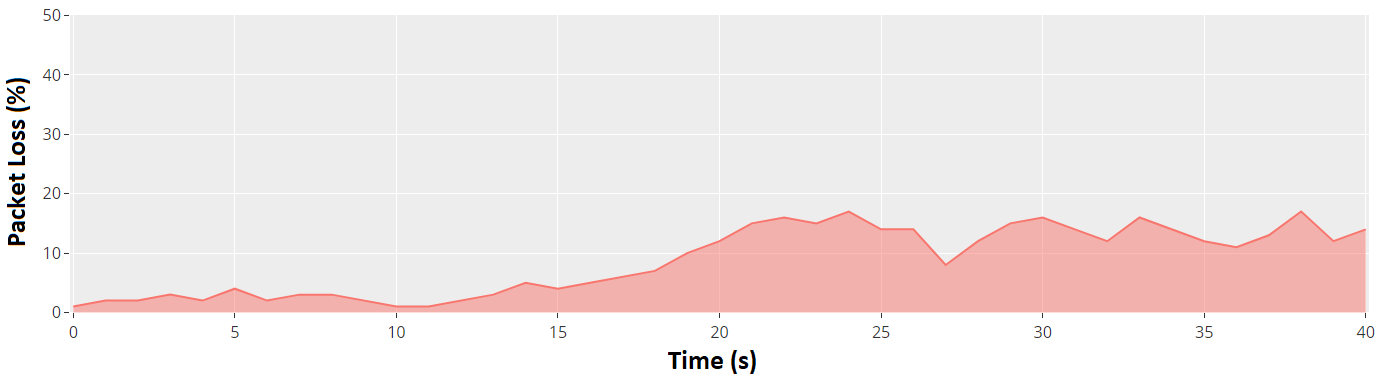}
    \caption{Packet loss analysis.}
    \label{fig:platooning_packetLoss}
\end{figure}

In both cases (increased and decreased frequency), the analysis found that the variable \emph{leader.acceleration} was a possible cause of the undesired behaviour, more specifically, its rate (i.e., the jolt, the third derivative of position). The sharp acceleration, followed by a deceleration from the leading vehicle, triggers a high frequency of packet transmission for a long time, which causes channel congestion. A congested channel increases the rate of packet loss, which measures the number of packets that do not arrive to the destination over the total number of packets that were sent. When this happens, the vehicles do not receive information in time, which increases the likelihood of collisions, as well as going out of range. Figure~\ref{fig:platooning_packetLoss} shows an analysis of the packet loss rate and how it grows to high levels as the simulation goes on.

\begin{figure}[!h]
    \centering
    \includegraphics[width=0.8\linewidth]{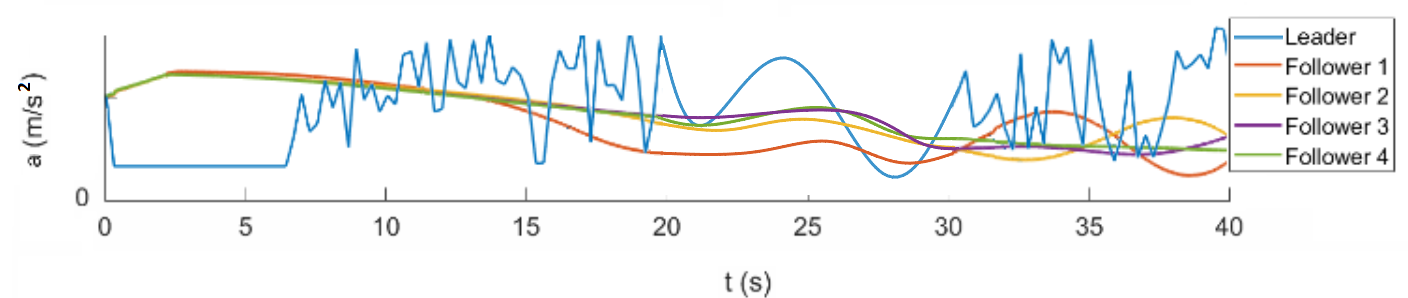}
    \caption{Overriding the leader trajectory.}
    \label{fig:platooning_causality}
\end{figure}

By overriding the leader acceleration, we decrease the acceleration rate, which in turn results in a smoother response from the followers, and thus prevents the hazards from occurring. Figure~\ref{fig:platooning_causality} shows the trajectory overriding in the $[20,30)$ time interval. 

Given that high channel congestion creates a hazard when vehicles are travelling at high speed, we would like to avoid that. As a way to correct our model, we have made adjustments such that whenever a high value for packet loss and channel congestion is detected, the model limits the acceleration rate until packet loss decreases to acceptable levels.


\subsection{Benchmarks}
\label{sec:benchmarks}

Here, we discuss the result of an experiment using CPS benchmarks found in the literature\footnote{https://cps-vo.org/group/ARCH/benchmarks} to assess and report on the scalability of our strategy. The benchmarks that are presented here can be used to assess different aspects of CPSs verification. All benchmarks were chosen such that they are scalable in different problem dimensions. 

The benchmarks originate from a variety of domains with the purpose of testing scalability with respect to the number of variables or locations. We make use of three benchmarks and each one assess a different scalability aspect.

This collection is organised by model and complexity. To ensure comparability of results, the specifications are unambiguous and formally described in hybrid automata~\cite{alur1993hybrid} which is a state-based modelling formalism for CPSs. One of the most constraining problem dimensions in hybrid systems verification are the number of continuous variables and the number of states (or locations) and transitions. All models are available with the HyConf tool\footnote{https://github.com/hlsa/HyConf}. The aim of this experiment is to evaluate the performance of the causal analysis. We measure its efficiency using 3 benchmarks systems. Moreover, each system comprises four variations that are incrementally complex (see Table \ref{tab:benchmark-systems}).

\begin{table}[!h]
\caption{Overview of the benchmark systems.}
\label{tab:benchmark-systems}
\centering
\begin{tabular}{|c|c|c|c|}  \hline
\textbf{System} & \textbf{Variables} & \textbf{Locations} & \textbf{Transitions} \\ \hline
Glycemic Control I & 11 & 6 & 10 \\ \hline
Glycemic Control II & 11 & 9 & 18 \\ \hline
Glycemic Control II & 11 & 12 & 32 \\ \hline
Glycemic Control IV & 11 & 15 & 56 \\ \hline
Filtered Oscillator I & 6 & 4 & 4 \\ \hline
Filtered Oscillator II & 10 & 4 & 4 \\ \hline
Filtered Oscillator III & 18 & 4 & 4 \\ \hline
Filtered Oscillator IV & 34 & 4 & 4 \\ \hline
Two-tank system I & 4 & 8 & 28 \\ \hline
Two-tank system II & 10 & 16 & 66 \\ \hline
Two-tank system III & 16 & 24 & 128 \\ \hline
Two-tank system IV & 22 & 34 & 296 \\ \hline
\end{tabular}
\end{table}

The Glycemic Control example is used to evaluate the increase in the number of locations and transitions whilst the number of variables remains constant. Conversely, the filtered oscillator is used to asses the increase in the number of variables whilst the structural components (i.e., locations and transitions) remain constant. Finally, the two-tank system is used to assess the increase in both the number of variables and number of locations and transitions, simultaneously. 

\subsubsection{Methodology}

The main goal of this study is the performance evaluation of our causal analysis. This experiment aims to verify whether the duration of the analysis grows linearly with the size of the subject system and the number of endogenous variables selected. Another motivation behind this study is the fact that there are few empirical and controlled experiments to evaluate the efficiency of causal analysis in general. 

\begin{itemize}
	\item \textbf{RQ1:} Does the duration of the causal analysis grow linearly with the number of selected endogenous variables? 
	\item \textbf{RQ2:} Does the duration of the causal analysis grow linearly with the number of variables in the subject system?
	\item \textbf{RQ3:} Does the duration of the causal analysis grow linearly with the number of locations and transitions in the subject system?
\end{itemize} 

We would like to note that it is possible to increase the number of endogenous variables without increasing the number of overall variables in the system by moving variables from the exogenous set of variables to the endogenous one. Conversely, it is also possible to increase the number of system variables without increasing the number of endogenous ones by adding them to the exogenous set.

We analyse the efficiency of our strategies using mutation analysis. The mutation operators used in this experiment were chosen based on a study on mutation operators for critical systems \cite{binh2012mutation}. We introduce five mutations for each system; all of the mutations change the value of variables in the systems. Then, we divide the executions in 3 stages. In each stage, we iteratively increment the number of endogenous variables considered in the search by a third of the total system variables. That is, in first stage, a third of the variables are considered as endogenous. In the second stage, we consider two thirds of the variables, and finally, in the third stage, every system variable is considered as endogenous. This choice is made randomly. We, then, collect the mean time to determine causes, which is followed by a statistical analysis of the results.

\subsubsection{Hypotheses}

In order to answer our research questions, we define the metric $TDC$, which represents the \textbf{t}ime to \textbf{d}etermine a \textbf{c}ause. Hypotheses A, B, and C aim to evaluate the research questions that have been explained previously, respectively. For this, null hypotheses are defined, which state that the duration of the causal analysis does not grow linearly with the increase in the complexity of the analysis. This experiment aims to refute such hypotheses. Thus, alternative hypotheses are also defined, which have a complementary role to the null hypotheses, and can be accepted in case its counterpart hypotheses are rejected. We define 6 hypotheses: A0 and A1 (null and alternative, respectively), checks whether the $TDC$ grows linearly with the number of endogenous variables. Analogously, hypotheses B0 and B1 (null and alternative, respectively) consider the number of variables in the systems. Lastly, C0 and C1 consider the number of locations and transitions in the system.

\begin{itemize}
	\item $H_{A0}$: The $TDC$ grows linearly with the number of selected endogenous variables. 
	\item $H_{A1}$: The $TDC$ does not grow linearly with the number of selected endogenous variables.
	\item $H_{B0}$: The $TDC$ grows linearly with the number of variables in the system. 
	\item $H_{B1}$: The $TDC$ does not grow linearly with the number of variables in the system.
	\item $H_{C0}$: The $TDC$ grows linearly with the number of locations and transitions in the system. 
	\item $H_{C1}$: The $TDC$ does not grow linearly with the number of locations and transitions in the system.
\end{itemize}

\subsubsection{Threats to validity}

Here we list the threats to validity that apply to this experiment. As
{\bf Internal Validity}, the mutation operators used in this experiment were chosen based on a study on mutation operators for safety critical systems \cite{binh2012mutation} and the number of inserted faults is decided manually. Concerning {\bf External Validity}, this experiment only considers 3 systems; we cannot generalise the outcome of this experiment for a general class of CPSs. Besides, since we introduced the faults ourselves, the mutants may also not represent real world faults. As {\bf Construct Validity},  the values of the mutants (i.e., the degree of the fault) have an impact on how difficult are for causes to be determined. Furthermore, the choice of endogenous variables has a big impact on the performance, as the cause may not be among the chosen variables.

\subsubsection{Results}

Table \ref{tab:benchmark-results} shows a summary of the results. For each row of Table~\ref{tab:benchmark-results} we show the respective system along with the number of variables, the number of locations and the number of transitions respectively. For instance, "Glycemic Control I (11 - 6 - 10)" represents the Glycemic Control I model that comprises 11 variables, 6 locations and 10 transitions. Furthermore, we depict the stage along with its respective number of endogenous variables and the mean time for the causality assessment. We would like to note that the time reported is the maximum time taken to analyse the system. We do not stop the search when a cause is found, even though there is only one cause by the experiment design. 

\begin{table}[]
\centering
\caption{Benchmark results.}
\label{tab:benchmark-results}
\resizebox{1\textwidth}{!}{%
\begin{tabular}{c|
>{\columncolor[HTML]{ECF4FF}}c |
>{\columncolor[HTML]{ECF4FF}}c |
>{\columncolor[HTML]{FFFFC7}}c |
>{\columncolor[HTML]{FFFFC7}}c |
>{\columncolor[HTML]{FFCE93}}c |
>{\columncolor[HTML]{FFCE93}}c |}
\cline{2-7}
 & \multicolumn{2}{c|}{\cellcolor[HTML]{ECF4FF}\textbf{Stage I}} & \multicolumn{2}{c|}{\cellcolor[HTML]{FFFFC7}\textbf{Stage II}} & \multicolumn{2}{c|}{\cellcolor[HTML]{FFCE93}\textbf{Stage III}} \\ \hline
\multicolumn{1}{|c|}{\textbf{System}} & \cellcolor[HTML]{ECF4FF}\textbf{\begin{tabular}[c]{@{}c@{}}End.\\ Vars.\end{tabular}} & \cellcolor[HTML]{ECF4FF}\textbf{\begin{tabular}[c]{@{}c@{}}Mean\\ Time\end{tabular}} & \cellcolor[HTML]{FFFFC7}\textbf{\begin{tabular}[c]{@{}c@{}}End.\\ Vars.\end{tabular}} & \cellcolor[HTML]{FFFFC7}\textbf{\begin{tabular}[c]{@{}c@{}}Mean\\ Time\end{tabular}} & \cellcolor[HTML]{FFCE93}\textbf{\begin{tabular}[c]{@{}c@{}}End.\\ Vars.\end{tabular}} & \cellcolor[HTML]{FFCE93}\textbf{\begin{tabular}[c]{@{}c@{}}Mean\\ Time\end{tabular}} \\ \hline
\multicolumn{1}{|c|}{Glycemic Control I (11 - 6 - 10)} & 4 & 6.7 min & 8 & 18.2 min & 11 & 41.6 min \\ \hline
\multicolumn{1}{|c|}{Glycemic Control II (11 - 9 - 18)} & 4 & 8.2 min & 8 & 18.9 min & 11 & 43.7 min \\ \hline
\multicolumn{1}{|c|}{Glycemic Control III (11 - 12 - 32)} & 4 & 9.2 min & 8 & 20.1 min & 11 & 44.9 min \\ \hline
\multicolumn{1}{|c|}{Glycemic Control IV (11 - 15 - 56)} & 4 & 10.4 min & 8 & 21.6 min & 11 & 46.3 min \\ \hline
\multicolumn{1}{|c|}{Filtered Oscillator I (6 - 4 - 4)} & 2 & 3.7 min & 4 & 4.8 min & 6 & 8.9 min \\ \hline
\multicolumn{1}{|c|}{Filtered Oscillator II (10 - 4 - 4)} & 3 & 8.9 min & 7 & 14.8 min & 10 & 23.5 min \\ \hline
\multicolumn{1}{|c|}{Filtered Oscillator III (18 - 4 - 4)} & 6 & 24.2 min & 12 & 50.1 min & 18 & 107.8 min \\ \hline
\multicolumn{1}{|c|}{Filtered Oscillator IV (34 - 4 - 4)} & 11 & 53.9 min & 23 & 96.4 min & 34 & 192.8 min \\ \hline
\multicolumn{1}{|c|}{Tank system I (4 - 8 - 28)} & 2 & 4.4 min & 3 & 10.2 min & 4 & 22.6 min \\ \hline
\multicolumn{1}{|c|}{Tank system II (10 - 16 - 66)} & 4 & 13.1 min & 7 & 28.7 min & 10 & 89.7 min \\ \hline
\multicolumn{1}{|c|}{Tank system III (16 - 24 - 128)} & 5 & 34.2 min & 11 & 71.3 min & 16 & 156.4 min \\ \hline
\multicolumn{1}{|c|}{Tank system IV (22 - 34 - 296)} & 7 & 58.2 min & 15 & 132.1 min & 22 & 253.5 min \\ \hline
\end{tabular}%
}
\end{table}

The algorithm's time complexity seems to increase not linearly with the number of selected endogenous variables but linearly with the number of variables and locations in the automaton. The time reported in Table~\ref{tab:benchmark-results} is displayed in minutes and considers both the search and simulation times. For statistical significance, each time reported is computed as an average of 5 executions. In each of these five executions, the selected endogenous variables ("End. Vars.") remain the same. 


We use the Glycemic Control model to study the effects of increasing locations and transitions but keeping the total number of variables constant. For each variation (I through IV), the results indicate a linear growth in the time taken, due to the increase in operations contained within each added location. For instance, we double the number of locations from 6 to 12 (in Glycemic Control I and Glycemic Control III, respectively) but the time taken to search and assess causality does not double but increases steadily. However, when doubling the number of selected endogenous variables (from 4 to 8), the time taken increases considerably. See, for instance, Glycemic Control I, where the mean time goes from 6.7 minutes to 18.2 minutes. 

The filtered oscillator model is used to assess the growth in the number of total variables in the system whilst the number of discrete locations and transitions remain the same. Similarly to the Glycemic Control model, a growth in the number of variables (and, with it, the number of operations) increases the search time linearly. For instance, when the number of selected endogenous variables remain the same, we notice only a modest increase in the amount of time taken when our strategy is applied to Filtered Oscillator II (10 system variables) compared to Filtered Oscillator III (18 system variables). 

The findings obtained by analysing the previous models, can also be observed with the tank system. This time, both the total number of variables and the number of locations and transitions are increased and the time taken increases linearly if the number of selected endogenous variables remains the same. The tank system can be used to analyse all three metrics of this benchmark. 

The observations of this experiment indicate that increasing the complexity of the subject systems increases the time taken linearly. However, increasing the number of selected endogenous variables has a much greater impact on the overall efficiency of the strategy. Thus, we refute hypothesis $H_{A0}$, however we accept hypotheses $H_{B0}$ and $H_{C0}$.

Finally, we consider how Halpern has calculated the complexity of determining causality as an NP-complete problem in terms of complexity\cite{halpern2016actual}. Note that, in Halpern’s calculations, the number of exogenous variables does not play a role in the complexity as they are all grouped into the context (u). What plays a role is the number of variables in the cause and the number of endogenous variables (particularly the size of the sets X, W, and Z - all subsets of the endogenous set). Thus, even though we have observed linear growth in some circumstances, this was only observed when the overall size of the system increased but the number of endogenous variables remained the same (i.e., the additional variables were all added to the set of exogenous ones). When the number of endogenous variables increases, then the time increases non-linearly. Moreover, we emphasise the limitations regarding the choice of endogenous variables and the maximum number of variables in a cause that were applied to our strategy in order to achieve these numbers.

\section{Conclusions}
\label{sec:conclusion}

Causal analysis is an essential ingredient of counterexample analysis conducted after verification techniques, providing an effective technique to isolate and eventually remove hazards in the design and failures observed in the verification process.

In this work, we propose a formal theory and a practical analysis technique for assessing causality in continuous systems. We attained the following results: 1) we extended an existing theory of actual causality \cite{halpern2005causes} to cope with cyber-physical systems 2) we developed a process to apply the theory of causality and developed and integrated the algorithms in our tool (HyConf \cite{araujo2019multiobjective}), and 3) we applied the developed technique to two case studies to evaluate its effectiveness.

As for future work, in the case where multiple causes are provided, we aim to rank the causes based on a weighted criteria (e.g., cost or responsibility \cite{chockler2004responsibility}). Moreover, the notation for describing the effects (i.e,  $\Phi$) should be extended to consider the temporal quantifiers of Signal Temporal Logic \cite{maler2004monitoring}. Furthermore, there is an important discussion to be had with respect to the use of (good) causal models. In our strategy, causal models will determine whether an appropriate cause can be found and, thus, the choice of causal model is of high importance. An option is to obtain and using input-output relationship descriptions from hybrid system models \cite{mitsch2016modelplex}, which can assist the user with building relevant causal models.

Currently, concerning the theory, we allow R to be any trajectory vector that obeys the signal types, but without considering any further constraints of the specification besides maximum and minimum values for variables. This may result in causes that do not obey specified dynamics. We aim to address this issue as future work by, during the search, limiting the derivatives to reasonable boundaries obtained by, for instance, exploring hybrid automata \cite{alur1993hybrid} models that can be fed as input.

Lastly, despite our measures to address performance-related issues, we believe it is possible to further improve efficiency of the tool. We aim to do this by studying the effect of different search strategies as well as analysing the impact of the length of the trajectories involved in the causal analysis (e.g., the one that contains the fault and the ones that contain the causes).

\begin{acks}
Hugo Araujo and Mohammad Reza Mousavi have been partially supported by the UKRI \resizebox{1\textwidth}{!}{
Trustworthy Autonomous Systems Node in Verifiability, Grant Award Reference EP/V026801/2.}
\end{acks}



\bibliography{references}
\bibliographystyle{plain}

\newpage

\begin{appendix}

\section{List of operators}
\label{ap:operators}

$\quad X$ \hfill Variable.

$\vec{X}$ \hfill List/Set of variables.

$x$ \hfill Trajectory.

$\vec{x}$ \hfill List/Set of trajectories.

$x \downarrow_{\vec{X}}$ \hfill Trajectory restricted to the set of variables $\vec{X}$.

$x_{[i,j)}$ \hfill Trajectory slice.

$x =_{[i,j)} y$ \hfill  The trajectories x and y are equals when restricted to the set $\vec{Z}$ and within [i,j).


$c \mapsfrom \vec{x}$ \hfill The values in the list of trajectory $\vec{x}$ are equals to the values in $c$.

$M_{\vec{X} \leftarrow \vec{x}}$ \hfill Filtering the trajectories of the variables in $\vec{X}$ with the trajectories in $\vec{x}$.

\section{Soundness}
\label{sec:soundness}

In order to discuss the soundness of our implementation of the definition of cause, we provide a one-to-one compositional mapping between the constructs used in  Algorithm~\ref{alg:causalityAlgorithm} and the concepts introduced in Definition~\ref{def:causeCPS}. Firstly, we provide Proposition~\ref{prop:cequalsxX} that makes the mapping easier to understand.

\begin{proposition}
\label{prop:cequalsxX}
In our implementation, $(M,u) \models (c \mapsfrom \vec{x})$ always holds by construction.
\end{proposition}

\begin{proof}
In our implementation, the causal model $M$ is built independently of a system execution. However, once a trajectory ($c$) that leads to a failure ($\Phi$) is found, the trajectory that corresponds to the exogenous variables ($u$) is built by applying the projection ($c \downarrow_{\mathcal{U}} $). Thus ($M,u$) is always built around $c$. Furthermore, the list of trajectory slices in the cause ($\vec{x}_X$) is, by construction, taken from $c \downarrow_{\mathcal{V}}$. Thus, Proposition~\ref{prop:cequalsxX} follows.
\end{proof}

In what follows, we provide a one-to-one compositional mapping of Definition~\ref{def:causeCPS} into Algorithm~\ref{alg:causalityAlgorithm} (see Table~\ref{tab:map-alg-def}) and explain it below. For that, we rely on the assumption that the inputs (generated by our search heuristic in Algorithm~\ref{alg:causalitySearchAlgorithm}) are generated appropriately from the models and that the causal model conforms with the system. In the preamble for the definition of cause, a causal model $M$, a trajectory $c$, and a conjunction of Boolean predicates $\Phi$ are given; analogously, they are given as input for our algorithm. Furthermore, the definition requires a list of variables $\vec{X} \in \mathcal{V}$ that is taken from the causal model in the search algorithm. Lastly, the theory also calls for a list of trajectory slices $\vec{x}$. In both the algorithm and the definition, to determine cause, three clauses must be satisfied (4th row of Table~\ref{tab:map-alg-def}). 

The mapping of clause AC1 is straightforward. By construction (in Algorithm~\ref{alg:causalitySearchAlgorithm}), $\vec{x}$ is taken from $c$ and, thus, it suffices for Algorithm~\ref{alg:causalityAlgorithm} to only evaluate whether $\Phi$ holds in the causal model. As for AC2, the set $\vec{W}$ is also given as input. An alternative trajectory is the result of overriding  $c$ with the slices $\vec{x}^\prime$ and $\vec{w}$, which is done via the model update (9th row). This is used to check that $\Phi$ does not hold in this updated model. Lastly, AC3 checks minimality by searching for causes that are subset of the prospective one. This new search adds trajectories to $\mathcal{R}$, which then confirms soundness as, according to Definition~\ref{def:causeCPS}, we need to check every $\vec{w} \subseteq \mathcal{R} \cap \mathrm{Trajs}(\mathcal{W})$. 

\begin{longtable}{|m{0.32\textwidth}|m{0.63\textwidth}|}
\caption{Mapping between Algorithm~\ref{alg:causalityAlgorithm} and Definition~\ref{def:causeCPS}.}
\label{tab:map-alg-def}\\

\hline
Given a causal model $M = \langle \langle \mathcal{U}, \mathcal{V}, \mathcal{R} \rangle, \mathcal{F} \rangle$ & \includegraphics[scale=0.29]{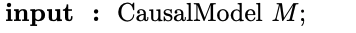} \\ \hline

a setting $u \in \mathit{Trajs}(\mathcal{U})$, a trajectory $c : \mathit{Trajs}(\mathcal{U} \cup \mathcal{V})$ such that $c\downarrow_{\mathcal{U}} =_{\mathrm{dom}(u)} u$ & 
\includegraphics[scale=0.29]{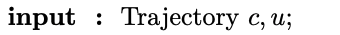}
\\ \hline

a list of variables $\vec{X}$,  & \includegraphics[scale=0.29]{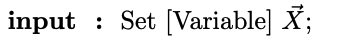} \\ \hline

and a set of trajectory slices $\vec{x} = \{x \mid x \in \mathit{Trajs}(\mathcal{V})\}$, & \includegraphics[scale=0.29]{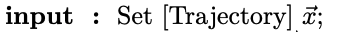} \\ \hline

then $(c \mapsfrom \vec{x})$ is a cause of $\Phi$ in $(M,u)$ when the following three conditions hold: & \includegraphics[scale=0.29]{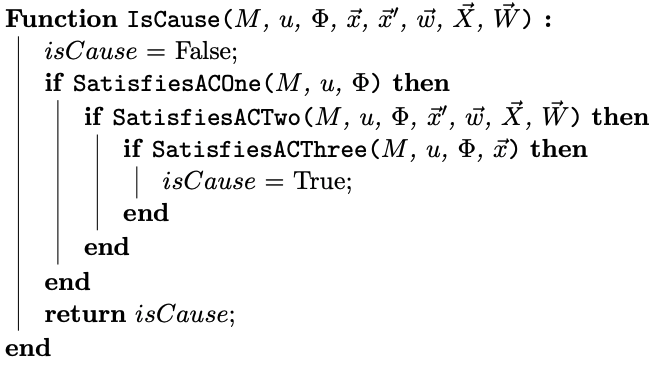}\\ \hline

AC1. $(M,u) \models (c \mapsfrom \vec{x}) \wedge \Phi$ & \includegraphics[scale=0.29]{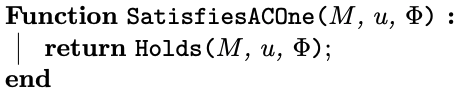}\\ \hline

AC2. There exists a set of variables $\vec{W} \subset \mathcal{V}$ &  \includegraphics[scale=0.29]{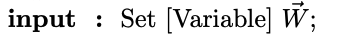} \\ \hline 

and two sets of trajectories $\vec{x}^\prime \in \mathit{Alts}(\vec{x})$, and $\vec{w} \subseteq \mathit{Trajs}(\vec{W}) \cap \mathcal{R}$ such that if $(M,u) \models (c \mapsfrom \vec{w})$, then: & \includegraphics[scale=0.29]{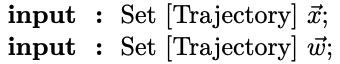} \\ \hline


$(M_{\vec{X} \leftarrow \vec{x}^\prime, \vec{W} \leftarrow \vec{w}}, u) \models \neg \Phi$ & \includegraphics[scale=0.29]{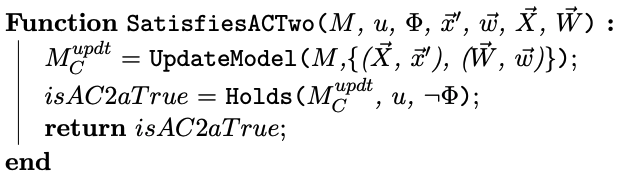} \\ \hline

AC3. There is no strict subset of $\vec{x}$ that satisfies AC1 and AC2 & \includegraphics[scale=0.27]{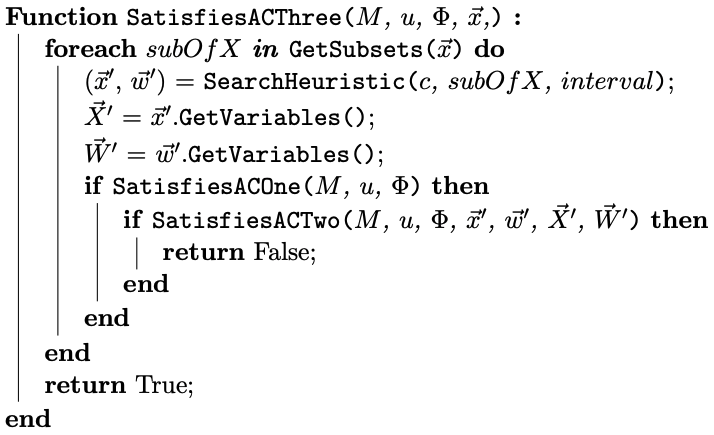}\\ \hline

\end{longtable}

The computational complexity of determining a cause is considered intractable (NP-complete for some cases~\cite{halpern2016actual}). And it is due to this reason that we apply some limitations to our strategy (as highlighted in Sections~\ref{ssec:process} and~\ref{ssec:verification}) in order to reach a verdict (e.g., the number of endogenous variables and a maximum number on interactions between variables during the search). Hence, our algorithm only provides an approximation to the solution of the actual problem. Since we impose finite limits to the number of chosen variables, the granularity and number of slices, and the number of interactions between variables (up to three-wise) when searching, we guarantee that the program will terminate. However, due to the problem's intractability and the pseudo-random nature of the search, we do not guarantee that our approach will find a cause. Our approach is not exhaustive but we have shown via the mapping that our approach is sound with respect to the discretised version of the problem. That is, our approach finds actual causes (with respect to our theory) when one considers the necessary discretisation steps (in which we approximate a strictly continuous problem into a discrete one), which are needed for simulation of CPSs in physical machines.

\end{appendix}

\ifdefined\CHANGES
\printindex[changes]
\fi

\end{document}